\documentclass[fleqn,usenatbib]{mnras}

\usepackage{newtxtext,newtxmath}

\usepackage[T1]{fontenc}
\usepackage{ae,aecompl}
\usepackage{multirow}

\newcommand{\microJybm}{\,\mathrm{\mu Jy\,beam^{-1}}}
\newcommand{\microJy}{\,\mathrm{\mu Jy}}

\usepackage{etoolbox}
\makeatletter
\patchcmd\@combinedblfloats{\box\@outputbox}{\unvbox\@outputbox}{}{\errmessage{\noexpand patch failed}}
\makeatother
\usepackage{graphicx}	
\usepackage{amsmath}	
\usepackage{amssymb}	
\usepackage{multirow}
\usepackage{natbib}
\usepackage{siunitx}
\usepackage{pdflscape}

\title[The variable extragalactic radio sky]{An insight into the extragalactic transient and variable microJy radio sky across multiple decades}

\author[J. F. Radcliffe et al.]{Jack F. Radcliffe,$^{1,2,3,4}$\thanks{E-mail: jack.radcliffe@manchester.ac.uk}
Robert. J. Beswick,$^{2}$
A.~P. Thomson,$^{2}$
Michael A. Garrett,$^{2,5}$
\newauthor
Peter D. Barthel,$^{1}$ and Thomas W. B. Muxlow$^{2}$
\\
$^{1}$Kapteyn Astronomical Institute, University of Groningen, 9747 AD Groningen, The Netherlands\\
$^{2}$Jodrell Bank Centre for Astrophysics, School of Physics \& Astronomy, The University of Manchester, Alan Turing Building, Oxford Road, Manchester M13 9PL, UK\\
$^{3}$ASTRON, the Netherlands Institute for Radio Astronomy, Postbus 2, 7990 AA Dwingeloo, The Netherlands\\
$^{4}$Department of Physics, University of Pretoria, Lynnwood Road, Hatfield, Pretoria, 0083, South Africa\\
$^{5}$Leiden Observatory, Leiden University, PO Box 9513, 2300 RA Leiden, The Netherlands
}

\date{Accepted XXX. Received YYY; in original form ZZZ}

\pubyear{2018}

\begin{document}
\label{firstpage}
\pagerange{\pageref{firstpage}--\pageref{lastpage}}
\maketitle

\begin{abstract}
The mJy variable extragalactic radio sky is known to be broadly non-changing with approximately $3\%$ of persistent radio sources exhibiting variability which is largely AGN-related. In the faint (<mJy) flux density regime, it is widely accepted that the radio source population begins to change from AGN dominated to star-formation dominated, together with an emergent radio-quiet AGN component. Very little is known about the variable source component in this sub-mJy regime. In this paper, we provide the first insight into the $\mu$Jy variable sky by performing a careful analysis using deep VLA data in the well studied GOODS-N field. Using five epochs spread across 22 years, we investigate approximately 480 radio sources finding 10 which show signs of variability. We attribute this variability to the presence of an AGN in these systems. We confirm and extend the results of previous surveys, finding that variability in the faint radio sky is rather modest with only $\leq$2\% of sources exhibiting significant variability between any two epochs. We find that 70\% of variable sources show variability on timescales of a few days whilst on longer decadal time-scales, the fraction of variable sources decreases to $<1\%$. This suggests that the radio variability peaks on shorter timescales as suggested by other studies. We find that 80\% of variable sources have VLBI counterparts, and we use multi-wavelength data to infer that these may well be core-dominated FR-I sources as postulated by wide-field VLBI surveys and semi-empirical simulations.

\end{abstract}
\begin{keywords}
radio continuum: transients -- galaxies: active -- techniques: interferometric
\end{keywords}
 


\section{Introduction}

Extremely deep radio observations of extragalactic fields have now revealed the nature of the persistent sub-100\,$\mathrm{\mu Jy}$ radio source population. It is generally agreed that at $\mathrm{\mu Jy}$ flux densities the radio source population is dominated by moderate to high redshift star-forming galaxies and radio-quiet active galactic nuclei (AGN). This is in contrast to the radio source counts at flux densities of $\sim$mJy and above, which are dominated by radio galaxies and quasars \citep[e.g.][and references therein]{Muxlow2005:sc,Padovani2015:eg,Smolcic2017:sc}.


Whilst deep field studies are providing a view of the ${\rm\mu Jy}$ radio sky in the spatial and spectral domains, at present comparatively little is known about their intrinsic variability in this faint flux density regime. Previous studies have shown that the variable and transient radio sky consists of a range of interesting astrophysical phenomena. Within our own galaxy, these can range from flaring stars and X-ray binaries \citep[e.g.][]{Thyagarajan2011:var,Williams2013:var} to novae and pulsars \citep[e.g.][]{Healy2017:var} whilst the transient and variable extragalactic radio sky includes AGN, tidal disruption events, supernovae, gamma ray bursts, fast radio bursts and the recently observed binary neutron star mergers \citep[e.g.][]{Berger2003:sn,Fong2014:grb,2007Sci...318..777L,Hallinan2017:gwt}. Of the slower transients (on timescales of minutes or longer), AGN are by far the most common extragalactic radio variable sources ($\gtrsim 60$ per deg$^{2}$ above 0.3 mJy) and are found to vary at all frequencies on time-scales from minutes to years \citep[e.g.][]{Dennett-thorpe2002,Lovell2008:scin,Ofek2011:var,Hodge2013:var,Mooley2013:var,Pietka2015:var,Mooley2016:var}. 

Many transient surveys are focused on relatively short-duration events, such as radio afterglows of gamma-ray bursts, however there have been few studies on the slowly-varying radio sky. From the few longer-term variability surveys that exist, the faint ($>0.5\,\mathrm{mJy}$) radio sky appears to be broadly non-variable, with only $\sim 3.6\%$ of sources showing variations over week to 1.5 year time-scales \citep{Mooley2016:var}. Over still longer time-scales (7-20 years), it was found that only $\sim$1\% of extragalactic sources in the mJy regime (excluding galactic transients) are variable \citep{Croft2010:var,Becker2010:var,Bannister2011b:var}.

\begin{table*}
	\centering
	\caption{A summary of the individual VLA/JVLA epochs utilised in this paper. The epochs correspond to the data that is compared on long timescales, typically 7-22 years whereas sub-epochs are those data used to establish any short term variability typically within a few days. This epoch and sub-epoch naming conventions used here are adopted throughout the paper. Note that the VLA observations were comprised of 24 observations spread over one month where the total observing time was approximately 47 hours on source.}
	\label{tab:observations}
	\begin{tabular}{ccccccc} 
\hline\hline
		Telescope & VLA Project Code & Epoch & Sub-epoch & Date (UT) & Frequency & rms \\
        & & & & & [GHz] & [$\rm\mu Jy\,beam^{-1}$] \\
		\hline
		VLA & AR0368 & VLA1996 & & 1996 Nov 01 - 1996 Dec 01 & 1.35-1.45 & 5.6 \\
        \hline
        \multirow{5}{*}{JVLA} & \multirow{5}{*}{TLOW0001} & \multirow{5}{*}{JVLA2011} & \multirow{2}{*}{JVLA2011 ep1} & 2011 Sep 09 12:57:47 - 17:56:57 & \multirow{5}{*}{1.00-2.03} & \multirow{5}{*}{2.80}  \\
         & & & & 2011 Sep 09 18:00:25 - 22:59:35  \\
        \cline{4-5}
         & & & \multirow{3}{*}{JVLA2011 ep2} & 2011 Sep 11 11:49:53 - 15:49:01   \\
         & & & & 2011 Sep 11 15:49:11 - 20:48:20   \\
         & & & & 2011 Sep 11-12 20:48:26 - 01:47:36   \\
        \hline
		\multirow{5}{*}{JVLA} & \multirow{5}{*}{18A-392} & \multirow{5}{*}{JVLA2018} & \multirow{2}{*}{JVLA2018 ep1} & 2018 May 09 06:27:27 - 08:25:28 &\multirow{5}{*}{1.00-2.03}& \multirow{5}{*}{3.98} \\ 
         & & &  & 2018 May 12 06:21:33 - 08:19:39  \\
          \cline{4-5}
         & & & \multirow{2}{*}{JVLA2018 ep2}  & 2018 May 15 03:24:20 - 05:22:26  \\
         & & &  & 2018 May 16 01:14:41 - 03:12:43  \\
           \cline{4-5}
         & & &  & 2018 May 19 01:23:49 - 03:34:10  \\
		\hline
	\end{tabular}
\end{table*}

However, these previous studies have a number of limitations. Long term (several years) variability surveys are reliant on archival data such as the FIRST/NVSS survey with limited sensitivity (1$\sigma$, 0.2 mJy) and thus are only able to probe the brighter end of the transient radio luminosity function. The next generation of radio interferometers, such as the Karl G. Jansky Array \citep[JVLA;][]{Perley2011:tel}, e-MERLIN, ASKAP \citep{Johnston2008:ask}, LOFAR \citep{vHaarlem2015:lo}, MeerKAT \citep{Booth2009:mer}, Apertif/WSRT \citep{Oosterloo2010:ap}, the Square Kilometer Array \citep[SKA;][]{Dewdney2009:ska} and the Next Generation VLA \citep[e.g.][]{Carilli2015:ng}, with their much improved survey speeds and snapshot sensitivity will be able to study the faint transient radio population across large areas of the sky.

The GOODS-N field has had multiple $\mu$Jy sensitivity observations at 1.4\,GHz spanning over 22 years using both the Very Large Array \citep[VLA;][]{Richards2000:hdf,Biggs2006:hdf,Morrison2010:hdf} and the recently upgraded JVLA \citep[][Radcliffe et al. in prep.]{Owen2018:hdf} with cadence times from a few days to decades\footnote{Note that in this paper, we use observations from before and after the VLA electronics upgrade hence the term JVLA will correspond to post-upgrade and the term VLA to pre-upgrade.}. In this paper, our goal is to use this unique data-set to investigate the $\rm\mu Jy$ variable sky over a 22-year period, probing the variable properties of over 400 hundred faint extragalactic sources.

This paper is organised as follows. In Section~\ref{Sec:Data} we present the observations, data reduction and imaging. We outline the steps used to define our variable sample in Section~\ref{Sec:variable_sample} and our results are presented in Section~\ref{Sec:results}. We discuss the implications of our results and compare to other variability surveys in Section~\ref{Sec:discussion}. Concluding remarks are given in Section~\ref{Sec:conclusion}.

Throughout this paper we adopt a spatially-flat 6-parameter $\mathrm{\Lambda CDM}$ cosmology with $H_0 = 67.8\pm0.9~\mathrm{km\,s^{-1}\,Mpc}$, $\Omega_{m} = 0.308\pm0.012$ and $\Omega_{\Lambda} =0.692\pm0.012$ \citep{Planck2016}.  For spectral index measurements, we use the convention $S_\nu \propto \nu^{\alpha}$ throughout, where $S_\nu$ is the radio integrated flux density and $\alpha$ is the intrinsic source spectral index.

\section{Data}\label{Sec:Data}
\begin{figure*}
	\includegraphics[width=\linewidth]{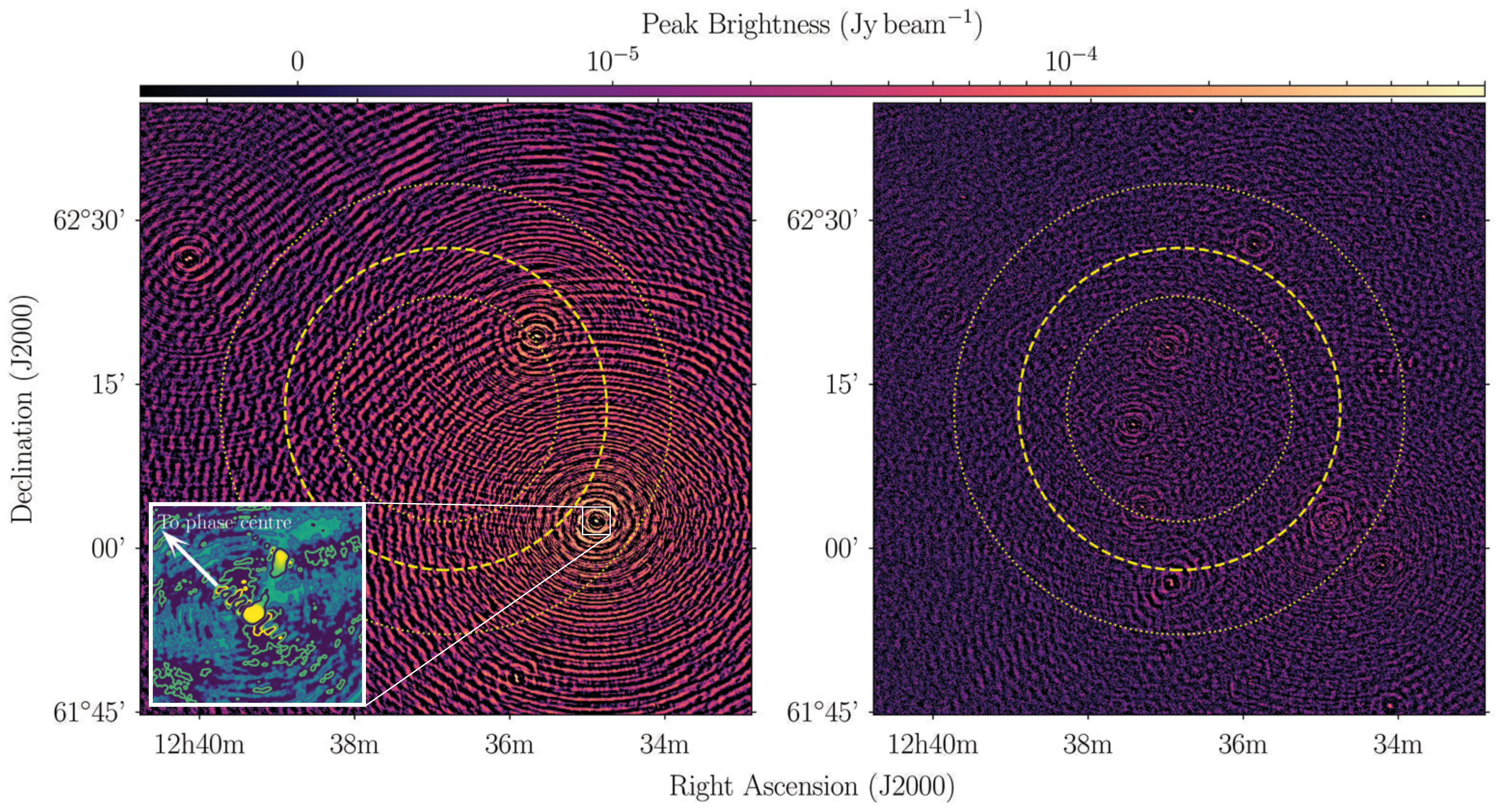}
    \caption{$1^{\circ}\times1^{\circ}$ image of the 1996 VLA data illustrating  the source removal routine used outlined in Section~\ref{SSec:source removal}. The yellow dashed circle corresponds to the HPBW of a VLA antenna at 1.4\,GHz ($\sim 29\arcmin.5$) while the dotted circles correspond to the VLA HPBW at 1\,GHz and 2\,GHz ($\sim 41\arcmin.2$ and $\sim 20\arcmin.6$ respectively). The symmetric logarithmic colour-scale is the same for both panels and ranges from ${\rm -5\,\mu Jy\,beam^{-1}}$ to ${\rm 0.8\,mJy\,beam^{-1}}$ \textit{Left Panel:} The data before the source removal process. Ripples from off-axis sources severely affect the central target field. Inset is a cut-out of S1 (J2000 12h35m38.044s +62d19m32.097s) illustrating the un-deconvolvable sidelobes present in the 1996 VLA data which are directed towards the phase centre. \textit{Right Panel:} The data post the removal of confusing sources revealing sources located in the centre of the field.}\label{Fig:peeling_results}
\end{figure*}
\subsection{VLA observations}

In November 1996, a total of 50 hours of VLA (A-configuration; VLA project code AR0368) 1.4\,GHz observations, centred within the GOODS-North region at J2000 RA $12^{\rm h}36^{\rm m}49.\!\!^{\rm s}4$ DEC $+62^{\circ}12^{\prime}58.\!\!^{\prime\prime}$0 were observed in a pseudo-continuum, spectral line mode. This was divided into two frequency sub-bands or spectral windows (spw), with central frequencies of 1.365\,GHz and 1.435\,GHz and a combined total bandwidth of 43.75\,MHz. Each spectral window comprised of 7$\times$3.125 MHz wide channels. For the purpose of this study we have re-analysed these data using modern absolute flux scales and advanced calibration techniques which were not available at the time of the original publication of \citet{Richards2000:hdf}.

The source J1313+6735 ($S_{\rm 1.4\,GHz}=2.4\,{\rm Jy}$) was used as a amplitude, phase and bandpass calibrator and was observed every 40 minutes. 3C286 was used as the flux density calibrator where the assumed flux densities of 15.328\,Jy and 14.947\,Jy at 1.385 and 1.435\,GHz respectively are derived from the \citet{PerleyButler2017:fs} absolute flux density scale. We discuss how differing flux scales between various observations are a vital consideration for any variability study in Section~\ref{SSec:Absolute_flux_scaling}. 

All calibration was conducted using the Common Astronomy Software Applications \citep[CASA;][]{Mcmullin2007:CASA}. The raw VLA data sets for the entire 50 hours were imported into a single $uv$ measurement set using \texttt{importvla}. Due to known issues with the old VLA MODCOMP control computers, the task \texttt{fixvis} was used to correctly re-calculate the $u,v$ and $w$ visibility coordinates. These data were then inspected and  Radio Frequency Interference (RFI) was removed using a combination of CASA tasks \texttt{flagdata}, \texttt{viewer} and \texttt{plotms} along with the \texttt{rfigui} viewer packaged with the \texttt{AOFlagger} software \citep{Offringa2012:rfi}. Once the strongest RFI was excised, a gain curve, describing the forward gain of each antenna, was derived using \texttt{gencal}. The absolute flux scale was set using an L-band model of 3C286 using \texttt{setjy} \citep{PerleyButler2017:fs}. Bandpass calibration was conducted on a per scan basis using calibrator, J1313+6735, and a complex gain calibration (amplitude and phase) was conducted on J1313+6735 and 3C286 with  five minute averaging interval using \texttt{gaincal}. This calibration table was used to bootstrap the flux density scale from 3C286 to J1313+6735. The integrated flux density of J1313+675 was found to be be 2.396\,Jy at 1.40\,GHz which is within 0.2\% of the quoted value of 2.40\,Jy.\footnote{\url{https://science.nrao.edu/facilities/vla/observing/callist}}

These calibration solutions were applied to the GOODS-N target field, which was split from the original data set. The data was reweighed to obtain the ideal sensitivity (using \texttt{statwt}), and any remaining RFI was excised from the data. Confusing sources were removed from the data following the procedures outlined in Section~\ref{SSec:source removal} and then self-calibration was performed on the target field (calibrating each polarisation separately) using multiple target sources, contained within a central 10\arcmin~diameter area, as a model. Due to the complexity of the sky model, self-calibration used an iterative strategy with solution intervals beginning at 20\,mins and culminating with 2\,min solution intervals for the final self-calibration cycle. No amplitude self calibration was conducted. 

\subsection{JVLA observations}

The JVLA (A-configuration) observed the GOODS-N field for a total of 38 hrs in between August and September 2011 (VLA project code TLOW0001; PI: F. Owen) and a further 10 hrs in May 2018 (VLA project code 18A-392; PI: J. Radcliffe). For the purposes of this study we only use 20 hours of the JVLA2011 data which provides two epochs with similar observing times (see Table\,\ref{tab:observations}). The observations have a total bandwidth of 1024\,MHz comprising of 16 spw with 64 channels per spw. 

The source J1313+6735 was used as a phase and bandpass calibrator and 3C286 was used as the flux calibrator. Initial flagging was conducted using \texttt{AOFlagger} \citep{Offringa2012:rfi}. The VLA CASA pipeline\footnote{\url{https://science.nrao.edu/facilities/vla/data-processing/pipeline}} was then used to conduct the flux density scaling and phase referencing. Some epochs required additional flagging (and re-calibration) after the initial pipeline calibration, especially if strong RFI had not been flagged adequately before or during calibration. In total, approximately 20-40\% of the data in each epoch was lost due to RFI. 


\subsection{Off-axis source removal}\label{SSec:source removal}

As has been noted in previous publications dealing with the GOODS-N field, there are significant issues with bright, confusing sources which limit the dynamic range of the central image. In order to deal with these sources, we were required to remove a number of sources from these data. Confusing sources were identified by generating a large, $2^\circ \times 2^\circ$, un-deconvolved image using \texttt{wsclean} \citep{Offringa2014:ws}.

The influence of off-axis confusing sources was more severe in the old VLA observations. The pre-WIDAR, VLA correlator had significant multiplicative off-axis errors when the continuum mode was used. These errors are correlated with the visibility strength resulting in un-deconvolvable side-lobes around the brightest sources which are orientated towards the phase centre (see the inset in Fig.~\ref{Fig:peeling_results}). This has been previously noted by \citet{Richards2000:hdf}, \citet{Biggs2006:hdf} and \citet{Morrison2010:hdf}. As a result, sources as far as 2$^\circ$ from the phase centre had to be carefully subtracted. For the JVLA data, these errors have been fixed with the upgrade to the WIDAR correlator. As a result, fewer sources needed to be peeled and the main source of confusion was due to the primary beam attenuation inducing direction-dependent errors upon bright sources near the primary beam half-power.

With confusing sources identified, each was removed individually using the following procedure:

\begin{enumerate}
	\item A frequency-dependent model of the source was generated using CASA task \texttt{tclean} and the multi-term multi frequency synthesis (MT-MFS) algorithm \citep{Rau2011:mtmfs}. Separate models were generated for each of the circular polarisations due of the significant beam squint of the VLA antennas.
	\item The model generated was then Fourier transformed and gridded in order to generate model visibilities (using task \texttt{ft}).
	\item Calibration corrections, derived by comparing the observed visibilities to the model visibilities and solving per antenna, were generated using \texttt{gaincal}. The observed visibilities were adjusted by these corrections using task \texttt{applycal}.
	\item Steps (i)-(iii) were repeated until the side-lobe structure around the confusing source was reduced.
	\item The final self-calibration model visibilities were subtracted from the observed visibilities using task \texttt{uvsub}. This removed the source and the side-lobe structure from the observed visibilities.
	\item The observed visibilities now have the correct phase and amplitude solutions at the location of the confusing source rather than the pointing centre. Therefore, we must revert our complex gains to the centre of the target field. To do this, the amplitude and phase corrections contained in final calibration table were inverted using the contributed CASA task \texttt{invgain} \citep{hales2016:inv}. 
	\item These inverted corrections were applied (using task \texttt{applycal}) to adjust the observed visibilities to be correct for the point centre again.
	\item Finally, steps (i)-(vii) were repeated for the other confusing sources.
\end{enumerate}

The results of this process are presented in Fig.~\ref{Fig:peeling_results}. This removal method is identical to the `peeling' method commonly used in radio interferometric data processing \citep[e.g.][]{OwenMorrison2008:sw,Intema2009:peel} with the exception that we do not reinsert the self-calibrated source model into the visibilities after step (viii). This meant that we were not required to deconvolve these sources when generating the final images, thus reducing the image size required.

\subsection{Imaging}\label{SSec:imaging}

A large $35\arcmin\times35\arcmin$ image was generated for each data set using CASA task \texttt{tclean}. Imaging for both the JVLA and VLA used the \texttt{wprojectft} algorithm \citep{2008ISTSP...2..647C}. This algorithm accounts for the non-coplanar array term (known as the $w$ term) thus minimising the induced smearing away from the phase centre. For these VLA data, deconvolution was performed using the \texttt{clarkstokes} algorithm. A primary beam model for the VLA correction was derived using the CASA recipe \texttt{makePB} and then was divided through the image using \texttt{impbcor}.

For these JVLA data, the imaging method was slightly different. Due to the large fractional bandwidth ($\sim68\%$), deconvolution was performed using the multi-term multi-frequency synthesis (MT-MFS) technique which permits more accurate deconvolution by taking into account the frequency dependence of the sky model \citep{Rau2011:mtmfs}. In addition, this method allows the correction of the difference in reference frequencies\footnote{The reference frequency is the centre of the band and, when using MT-MFS, is where the wide-band image flux density should equal the source flux density (see Section~\ref{SSec:ref_freq})} utilising the in-band spectral indices to correct fluxes to that of the same frequency as the VLA data. This technique is tested and outlined in Section~\ref{SSec:ref_freq}. The images were corrected for the primary beam attenuation and primary beam induced spectral curvature using the \texttt{widebandpbcor} task. The final data sets, epochs and sensitivities of the various epochs are described in Table~\ref{tab:observations}. 

\section{Defining a variable sample}\label{Sec:variable_sample}

\begin{figure*}
	\includegraphics[width=0.95\linewidth]{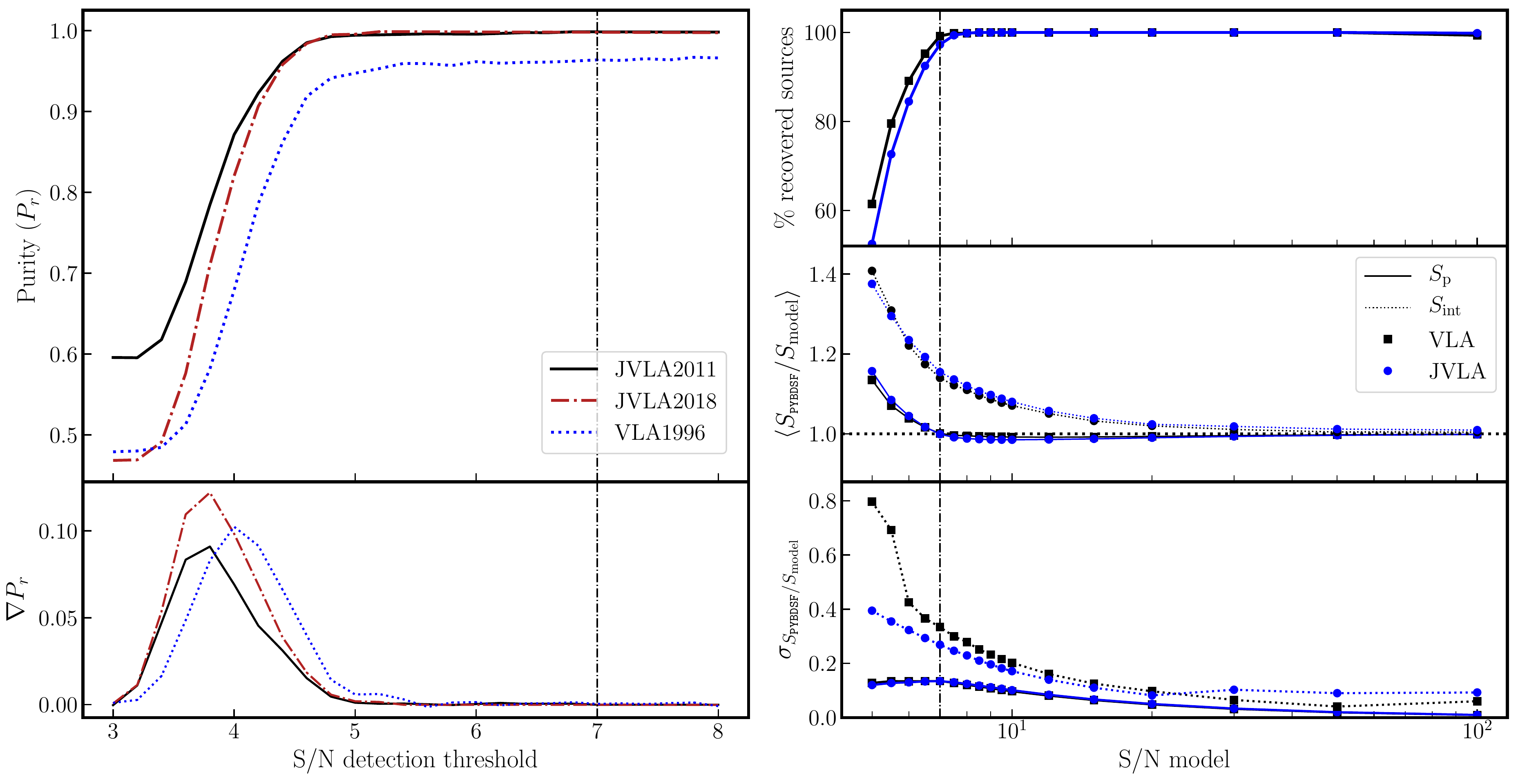} 
	\caption{Summary of the various source extraction tests performed using \texttt{PYBDSF}. \textit{Left Panel:} The purity, $P_r$ parameter against S/N detection threshold. The number of false detections in all epochs ($>3\%$) is large at $\mathrm{S/N} <5$. The parameter asymptotes for all epochs at around S/N thresholds of 5.5 as illustrated by the gradient of the purity parameter, $\nabla P_r$, in the bottom panel. \textit{Right Panel:} The results from the flux recovery simulations using \texttt{PYBDSF}. The top panel illustrates the percentage of model sources recovered by the source extractions software. The middle panel shows the average flux recovery performance while the bottom panel shows the typical 1$\sigma$ variance on the average recovered flux densities. The 7$\sigma$ detection threshold adopted in these analyses provides good peak brightness recovery with a typical error of around $12\%$ and recovers the majority of sources inserted $\sim 98\%$.}\label{Fig:catalog_tests}
\end{figure*}

With the images generated, we can now proceed onto the variability analysis. This section aims to summarise our selection method for defining the variable population while outlining and addressing the various systematic offsets which could induce artificial variability between the various epochs. These effects include the source extraction software, the reference frequency, the differing $uv$ coverage and the absolute flux scale. We shall address in turn each of these issues. Because we expect from previous studies that the number of variable sources will be low, systematics which can affect individual source flux densities (e.g. source extraction) are much more important to address than those which affect the source flux densities as a whole (e.g. absolute flux density scaling) as these can be modelled and corrected. 


\subsection{Source extraction}\label{SSec:source_extraction}

One of the most important systematics to address is the varying performance of the chosen source extraction routine. There are two main effects to consider, the first being the false detection rate and the second being the recovered flux densities of the extraction software. To identify sources we will use \texttt{PYBDSF} \citep{mohan2015:pybdsf}.

\subsubsection{False detection fraction}

In order to determine the appropriate signal-to-noise (S/N) threshold at which to allow {\sc PYBDSF} to identify and catalogue sources in our maps, we adopt the empirical `purity' parameter of \citet{Stach2018:pur}. We ran {\sc PYBDSF} on the real and inverted images (i.e. the real maps multiplied by $-1$) for each epoch with S/N thresholds between $3$-$8$, and we then evaluated the `purity parameter',

\begin{center}
  \begin{equation}
    P_{\rm r}=\frac{N_{p}-N_{n}}{N_{p}},
  \end{equation}
\end{center}

\noindent for each image, where $N_{p}$ is the number of sources detected above a given S/N limit in the real image and $N_{n}$ is the number of sources detected above the same S/N limit in the inverted image. For an idealised image comprising of Gaussian noise and a real source population, at an arbitrarily high S/N threshold the purity parameter should asymptote towards unity (i.e. $N_{p}>0$ and $N_{n}=0$). However, at lower S/N thresholds, we expect to detect a combination of genuine sources and spurious noise peaks in the real image, and only spurious noise peaks in the inverted image. 

As Figure~\ref{Fig:catalog_tests} illustrates, the purity parameter asymptotes towards unity around a S/N threshold of 5 for epochs observed with JVLA, whereas for the pre-upgrade VLA maps the purity parameter asymptotes to $\sim 0.97$. We determine that the VLA false-positive rate asymptotes to $\sim 3\%$ for ${\rm S/N}\gtrsim 5.5$. This is indicative of non-Gaussian noise properties in the VLA maps due to a combination of multiplicative baseline errors (which cause visible corrugations in the maps towards the pointing centre), and the more limited $uv$ coverage of the old array with respect to the post-2010 JVLA. Visual inspection of the inverted maps confirms this and these highly localised areas are excluded during subsequent cataloguing.

\subsubsection{Flux recovery}

In addition to ensuring that the S/N threshold reduces the number of false positives to a minimum, for variability studies, we also have to evaluate the source-extraction technique and any systematic biases to the recovered flux densities. One of these biases is the well-noted `flux-boosting' effect \citep{Jauncey1968,Vernstrom2016:con}. This causes the fitted flux densities, especially in the low S/N regime, to be systematically over-estimated due to a combination of confusion and instrumental noise. For deep radio surveys, this manifests itself as an over-estimation of the faint source counts. In the case of variability studies, this effect can be significant if the sources between the two epochs being compared have differing S/N ratios. The flux boosting effect is a function of S/N, therefore a source will be flux boosted in the less sensitive image resulting in artificial variability between epochs. The severity of this effect is also related to the choice of the source extraction routine. 


In these tests, we performed a hybrid approach which uses real observations (including residual calibration errors) with simulated sources injected. This was done separately for the VLA and JVLA due to the differing $uv$ coverage and therefore different noise characteristics. Before simulated sources were injected, a sky model was generated using \texttt{wsclean}, utilising the auto-masking routine to ensure that only real sources are removed and the noise characteristics are not changed. This model is Fourier transformed and subtracted from the visibilities and then the $uv$-data is imaged again to generate blank $7\mathrm{k}\times7\mathrm{k}$ pixel noise fields. These images are used to generate an r.m.s. map using \texttt{PYBDSF}. 

These model sources comprise of simple delta functions with a range of S/N ratios. These are convolved with the restoring beam and added into the noise fields. The positions of these sources are randomised across the field of view with the restriction that a source cannot be 10 pixels from another source in order to prevent blending complications. In addition, the periphery of the image ($\sim 10\%$ of the total image size) is avoided in order to reduce complications due to CLEAN aliasing. 
Each image is then catalogued using \texttt{PYBDSF} using a 5.5$\sigma$ detection threshold (as motivated by the purity tests). 

The results of these simulations are summarised in Figure~\ref{Fig:catalog_tests}. The peak brightnesses of the simulated sources are generally recovered by \texttt{PYBDSF} at around a S/N of 7. Below this threshold, the detected source flux distribution is censored where some sources which have their flux densities reduced by natural noise fluctuations are not detected. As a result, we see the flux-boosting effect which increases the average recovered flux densities in the low S/N regime. It is worth noting here that at intermediate S/N ratios ($\mathrm{S/N}=10$-$40$) the peak brightnesses are, on average, underestimated by $\sim$1\%. Whilst only small, this effect occurs due to the overestimation of the source size which reduces the fitted peak brightness while increasing the integrated flux densities. This is illustrated by the integrated flux densities being, on average, 5-10\% larger than the peak brightnesses in this S/N regime. In light of these conclusions, we adopt a conservative S/N threshold of $7\sigma$ for our detection threshold which provides a high reliability (as shown by the purity and sources recovered metrics) and adequate peak flux recovery (albeit with $\sim12\%$ errors) in this S/N regime.

\subsection{$uv$ coverage}\label{SSec:uv_coverage}

One potential source of induced variability could be caused by the differing $uv$-coverage of the various epochs. The multi-frequency synthesis imaging approach improves the sensitivity and fidelity of interferometric images by utilising the large bandwidths to fill the $uv$ plane. The large differences between the bandwidths of the VLA and JVLA observations therefore causes a mismatch in the $uv$ coverage. This effect is more significant on complex structures with differing radio power on multiple spatial scales, however, the majority of the radio sources in this field are not extended and subtend only 1-2 VLA beam widths. To address this issue, we excluded large extended sources (with classical radio jet structures) by masking them before cataloguing, and we did not consider those sources that required models comprising of multiple Gaussian components by \texttt{PYBDSF} in order to model their flux distribution. In addition, we only consider the peak brightnesses of sources as these radio emission mechanisms will have the same power on all spatial scales and are thus less sensitive to differences in the $uv$ coverage. It is worth noting that these cuts will inevitably bias our result as we may miss variable sources associated with extended objects. However such variability would be difficult to disentangle from artificial variability induced by the differing $uv$ coverage between the epochs.

\begin{figure}
	\includegraphics[width=\linewidth]{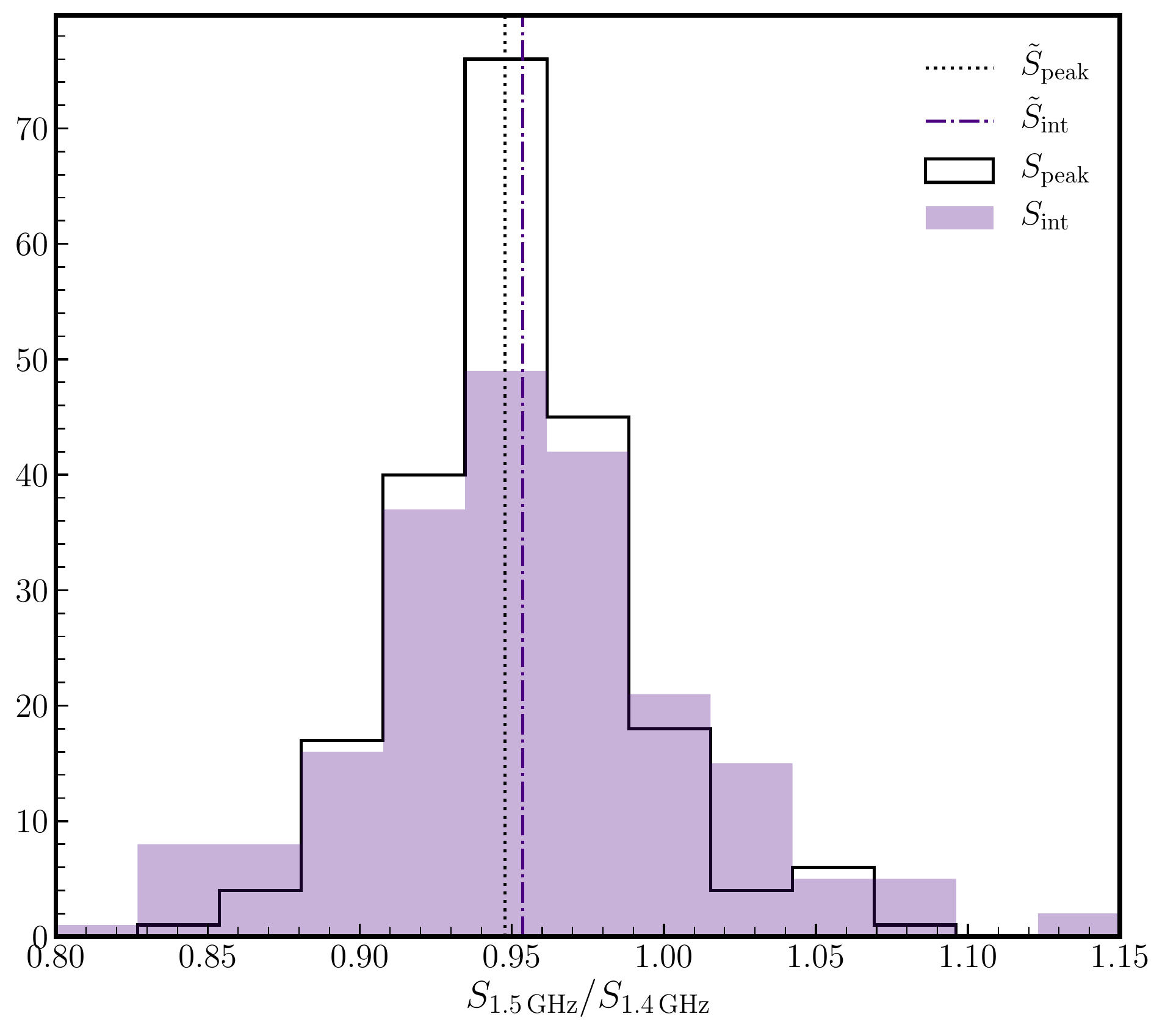}
    \caption{Comparison between the 1.4\,GHz and 1.52\,GHz fluxes of the JVLA data of epoch 5 (observed in 2018). The indigo filled and black edged histograms correspond to the ratio of peak brightnesses and integrated flux densities respectively. The corresponding dashed lines indicated the median of each distribution. The large scatter of both distributions are dominated by the intrinsic scatter in the individual source spectral indices.}	\label{Fig:ref_freq}
\end{figure}

\subsection{Reference frequency}\label{SSec:ref_freq}

The reference frequency is the frequency in which the wide-band flux densities in interferometric images equal the narrow band flux densities. This frequency $\nu_\mathrm{ref}$ is simply defined as the centre of the observing band,

\begin{equation}
\nu_\mathrm{ref} = \frac{\nu_\mathrm{high} - \nu_\mathrm{low}}{2}
\end{equation}

\noindent where $\nu_\mathrm{low}$ and $\nu_\mathrm{high}$ corresponds to the low and high edges of the observing bandwidth\footnote{see \url{https://casa.nrao.edu/casadocs/casa-5.6.0/imaging/synthesis-imaging/wide-band-imaging} for more information}. The intensity of an image is such that the wideband flux density is equal to the narrow-band source flux density at the reference frequency. Without taking into account the frequency dependence of the sky brightness distribution, this intensity is simply the weighted mean of the flux density across the bandwidth. This makes the returned fluxes susceptible to bias from the flagging of large frequency ranges. However, if MTMFS is used, the flux densities are fitted across the band and so the returned wide-band flux density will be closer to the real narrow-band flux density (Emonts private communication). This makes the returned flux densities more robust to any flagged data.

One possible source of induced variability is the differing reference frequencies between the VLA (1.4\,GHz) and JVLA (1.52\,GHz) epochs. Over the entire radio source sample, the differences between the two reference frequencies will manifest as a systematic flux scaling issue around the median spectral index of the source population. Indeed, \citet{Owen2018:hdf} finds a $\sim6\%$ systematic flux density decrease between the JVLA data used in these analysis and the \citet{Morrison2010:hdf} VLA observations which consistent with the difference in flux density expected between 1.4 and 1.52\,GHz with a mean spectral index in the expected range of $-0.7$ to $-0.8$. While the net effect of this can be fitted and removed, the large intrinsic scatter in the spectral index distribution \citep[$\alpha = -0.73\pm0.35$,][]{Smolcic2017:sc} means that a small fraction of sources with extreme deviations could significantly change the flux densities between 1.4 and 1.52\,GHz. 


For studies of the entire population as a whole, this is not significant, but in the case of variability where we expect a small number of variable sources, this effect could be significant on the final variable population identified. To correct for this, we utilise the MT-MFS algorithm when imaging the JVLA epochs to adjust the reference frequency to 1.4\,GHz. This algorithm uses the in-band source spectral index to re-scale each source to the appropriate flux density at 1.4\,GHz. To test that this was performing as expected, we used a single JVLA data set ( JVLA2018, observed in 2018). Two images covering an $18\arcmin\times15\arcmin$ were generated using CASA task \texttt{tclean} implementing the MT-MFS (with \texttt{nterms=2}) mode. For one of the images, the reference frequency for where the evaluation of the Taylor expansion, which takes into account the frequency dependence of the sky model during deconvolution, was set to 1.4\,GHz. Both of these images were primary beam corrected using the \texttt{widebandpbcor} task. These were then identically catalogued using \texttt{PYBDSF} with an arbitrary $8\sigma$ detection threshold (to reduce scatter induced by the fitting routine). As Fig.\,\ref{Fig:ref_freq} shows, the median difference between the peak brightnesses and integrated flux densities are 5.2\% and 4.7\% respectively. This corresponds to an median spectral index of $-0.67$ which is consistent with the literature \citep[e.g.][]{Condon1984,Kimball2008,Smolcic2017:sc}, thus validating this correction method in accounting for the reference frequency differences. 

The scatter in the distribution indicates that there is a non-negligible number of sources where this difference can cause 10-20\% amplitude adjustments. While this is not sufficient for outright classification using the metric used in this paper\footnote{A 20\% flux difference would only give a $V_s$ (Eqn.\,\ref{variability_selection}) of 1.414 assuming a typical 10\% error due to calibration and fitting}, the combination with other factors, such as residual calibration errors, could induce artificial variability in a small number of sources.   

Another possible artificial source of variability due to the reference frequency shifting correction could be due to the induced source spectral index causes by differing primary beam attenuation across the bandwidth. The spectral index, used by MTMFS to re-scale the flux densities to 1.4\,GHz, would be a linear combination of the source-intrinsic spectral index and a primary beam induced spectral index, thus causing an incorrect rescaling of the flux densities. The extra primary beam spectral index term, $\alpha_{\mathrm{E}}$, can be approximated by\footnote{see \url{https://casa.nrao.edu/casadocs/casa-5.6.0/imaging/synthesis-imaging/wide-band-imaging}}: 
	
\begin{equation}
\alpha_{\mathrm{E}}=-8 \ln (2)\left(\frac{\theta}{\theta_{0}}\right)^{2}\left(\frac{\nu}{\nu_\mathrm{ref}}\right)^{2}
\end{equation}

\noindent where $\theta_{0}$ is the primary beam FWHM at the reference frequency, $\theta$ is the distance from the pointing centre and $\nu$ is the observing frequency. Therefore, a flat spectrum source ($\alpha =0$) located 10\arcmin~from the pointing centre would have a additional primary beam induced spectral index of $\sim-0.89$. This would correspond to a shift in flux density of $\sim 7\%$ between the frequencies of 1.52\,GHz and 1.4\,GHz. To solve this, the wide-band primary beam correction, applied with CASA task {\tt widebandpbcor}, models and removes the primary beam induced spectral index, $\alpha_E$.

\subsection{Absolute flux scaling}\label{SSec:Absolute_flux_scaling}

Due to the use of the VLA pipeline, a different flux scale was used for the VLA and JVLA data during calibration. However, there is little difference between these flux density scales and any discrepancies are within the 3-5\% error \citep{PerleyButler2017:fs}. Indeed, the phase calibrator flux densities of all epochs agree fairly well and are all within 5\% of each other at 1.4\,GHz. These small offsets should be proportional to the source flux density and can therefore can be modelled and correction factors derived. 

To do this, we use robust linear regression (implemented in the \texttt{ODR scipy} Python package) to fit the peak fluxes between each pair of epochs in order to derive single epoch correction factors. Only sources with peak flux densities larger than 55$\mathrm{\mu Jy\,beam^{-1}}$ were considered because this corresponds to $10\sigma$ in the least sensitive epoch (VLA 1996). The more sensitive JVLA 2011 epoch 2 is used as the reference epoch. The scaling factors derived and applied to the peak brightnesses were 1.01, 1.04, 0.96 for the JVLA 2011 epoch 1, VLA 1996 and JVLA 2018 epochs respectively. 

\begin{figure*}
	\includegraphics[width=\linewidth]{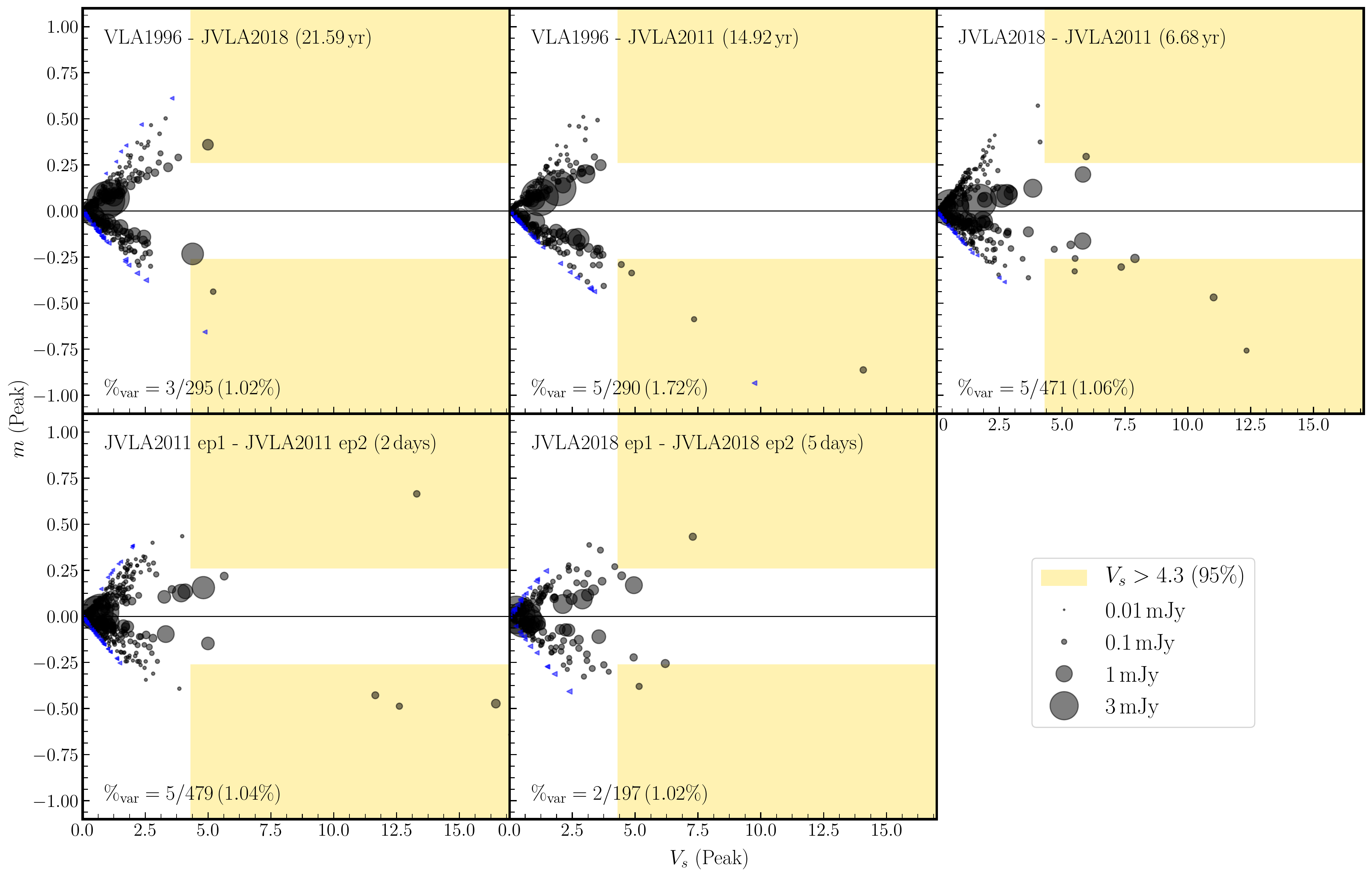}
	\caption{The two epoch comparison comparing the variability statistic $V_s$ to the modulation index, $m$. The black markers correspond to sources which are present in both epochs whilst the blue triangles are those sources which have upper limits in one epoch. The marker sizes are proportional to the mean peak brightness, $\langle S_p\rangle$. The gold region corresponds to the variability criteria where the variability must be statistically significant i.e. $V_s>4.3$ and it must have adjusted flux by 30\% (i.e. $|m| > 0.26$). The total number of variable sources is higher in this figure as some sources are classed as variables in more than one two epoch comparison. }\label{Fig:mod_index}
\end{figure*}

\subsection{Variability definition}\label{SSec:Variability_definition}

In order to identify variable and transient sources we adopt the formulation developed by \citet{Mooley2016:var}. We assume that our radio interferometric noise is Gaussian distributed \citep{Condon1998} and therefore we can compare the flux densities of a source, $S$, across two epochs (1,2) with the following equation:

\begin{equation}
\frac{\Delta S}{\sigma} = \frac{S_1 - S_2}{\sqrt{\sigma_1^2 + \sigma_2^2}}.
\end{equation}

\noindent where $\sigma$ is the measurement error. This comprises of the \texttt{PYBDSF} fitting error\footnote{This error is based upon the \citet{Condon1997} formulation and does not require the Monte-Carlo approach used in \texttt{PYBDSF} to evaluate errors as these are only needed for extended sources that are not considered here}, the in-band spectral index error ($<2\%$), and a surface brightness error to quantify slight differences in the $uv$ coverage, pointing errors, and primary beam correction. We assume a flux density error for the JVLA and VLA data of between 3-5\% \citep[][]{Morrison2010:hdf,Perley2013,PerleyButler2017:fs}.  %
The quantity, $\Delta S/\sigma$, is expected to follow the two-sided Student-t distribution and, utilising the same notation as \citet{Mooley2016:var}, we define the variability t-statistic, $V_s$ as:

\begin{equation}
\label{variability_selection}
V_s = \left|\frac{\Delta S}{\sigma}\right|
\end{equation}

\noindent The extent of a source's variability between two epochs can be quantified using the `modulation index', $m$, which is defined as,

\begin{equation}
m = 2 \frac{S_1-S_2}{S_1+S_2} = \frac{\Delta S}{\langle S\rangle},
\end{equation}
\noindent where $\langle S\rangle$ is the mean flux density of the source. Due to the expected low number of variable sources, we are more concerned with reliability over completeness. Therefore we adopt the rather conservative criterion for a variable source which is identical to that used by \citet{Mooley2016:var}. We define a source as variable if the variable t-statistic, $V_s$, is beyond the 95\% confidence interval ($V_s \geq 4.3$)\footnote{For two degrees of freedom, the 95\% confidence interval in the Student-t distribution corresponds to a Gaussian probability of more than $\pm 4\sigma$. For the Gaussian distribution, $4\sigma$ corresponds to a probability of about 1/16,000 (0.00625\%). The number of individual measurements in our variability analysis is 3464 which corresponds to 0.2165 false-positive variable sources. See \citet{Mooley2016:var} and p. 65-67 and Table C.8 of \citet{Bevington2003} for further details.} and its peak brightness changes by $\geq30\%$ (corresponding to $|m|>0.26$).

\subsection{Summary}

To summarise, the definitions and systematics outlined above leads to the following strategy being adopted in this paper, in order to identify variable and transient sources,

\begin{enumerate}
	\item Each epoch was searched for extended sources and these were identically masked in all epochs. 
	\item For each epoch, sources were extracted using \texttt{PYBDSF} using a S/N threshold of 7$\sigma$ based upon the false detection and flux recovery tests outlined in Section~\ref{SSec:source_extraction}. The total number of radio sources considered is around 250 to 480 due to the differing sensitivities of the epochs.
	\item Flux scaling issues due to small errors in the absolute flux scaling and differences in restoring beams are corrected using the routine outlined in Section~\ref{SSec:Absolute_flux_scaling}.
	\item These catalogues were cross-matched and combined using a 1\arcsec~search radius. Sources not present in some epochs have upper limits derived which correspond to 7 $\times$ the local r.m.s. 
	\item Using the definitions outlined in Section~\ref{SSec:Variability_definition}, $V_s$ and $m$ were derived for all two epoch combinations and variable candidates are identified if $V_s\geq4.3$ and $|m|\geq0.26$.
	\item Sources only present in one of the two epochs being compared have upper limits derived which correspond to the 7$\sigma$ detection threshold at the source position where $\sigma$ is derived using the r.m.s. maps generated by \texttt{PYBDSF}. If the detection threshold is smaller than the peak brightness then $V_s$ and $m$ are calculated, using the detection threshold as the peak brightness, otherwise this source is not considered in the two epoch comparison.
	\item Next we separate those sources into transient and variable sources. A source is classed as being variable if the source satisfies the variability criterion and is detected in multiple epochs while a transient source must satisfy the variability criterion but is only detected in a single epoch. 
	\item Finally, variable and transient candidates are checked manually to see if other factors such as nearby calibration errors or poor source extraction fitting could have artificially induced the variability. 
\end{enumerate}

\section{Results}\label{Sec:results}
\subsection{Variable sources}
The aim of this paper is to investigate the long-term $\mathrm{\mu Jy}$ radio sky. In order to achieve this, we split our JVLA epochs into two sub-epochs separated by short day-week timescales. This is done to establish whether any long term variability behaviour is artificially induced by any short term variability. In light of this, we establish two definitions corresponding to those sources with long and short term variability respectively. The long-term variable candidates are those which satisfy the variability criteria in any two-epoch comparison between the VLA1996, JVLA2011, and JVLA2018 epochs, while the short-term variables are identified as those which satisfy the variability criteria for any two sub-epoch comparisons (i.e. JVLA2011 ep1 and JVLA2011 ep2, JVLA2018 ep1 and JVLA2018 ep2). In Figure~\ref{Fig:mod_index} we show the variability index, $V_s$ against the modulation index, $m$, for all two-epoch and sub-epoch comparisons considered here. Sources which are located in the gold shaded areas satisfy the variability criteria ($V_s >4.3$ and $|m|>0.26$). The top row corresponds to the long-term variables while the bottom row corresponds to the short-term variables. In total, there are around 479 unique sources being considered in these analyses. The number of sources varies between approximately 200 and 480 due to the differing sensitivities of the epochs.

Initial investigations revealed 15 unique variable sources of which 5 were excluded after manual inspection. These were excluded due to a combination of poor source extraction, extended structure, residual calibration errors and smearing effects. In total, therefore we found 10 unique sources based upon the epoch and sub-epoch comparisons. The epoch-averaged peak flux densities ranging from 86 to 419 $\mathrm{\mu Jy\,beam^{-1}}$ with an average flux of 173 $\mathrm{\mu Jy\,beam^{-1}}$. The paucity of low flux density sources $<100\mathrm{\mu Jy\,beam^{-1}}$ may be indicative of the change in the radio properties of galaxies being primarily driven by AGN to those driven by star-formation processes, however we note that the variability statistic $V_s$ is biased towards large S/N, because the fitting error increases with decreasing S/N. 

For the candidate long-term variables, around 1\mbox{-}1.8\% of the persistent radio sources exhibit significant variability. Note that for the rest of this paper, the term `variable fraction' indicates the percentage of sources which exhibits variability within the total persistent radio population. There is no clear correlation between the cadence time and the variable fraction, however it is worth noting that the JVLA2018 and JVLA2011 two-epoch comparison is more sensitive ($\sim 30.5\,\mathrm{\mu  Jy\,beam^{-1}}$ detection threshold) hence the number of persistent sources compared differs. If we restrict the JVLA2011 and JVLA2018 epochs to the VLA1996 detection threshold, then we achieve a similar variable fraction of 1.4\%. In total, we find eight sources which are classed as long-term variables. Five of the eight sources are classed as variable on multiple two-epoch comparisons, while two show variability between all epochs. 

We classify $\sim1\%$ of our sample as candidate short-term variables, based upon measurements of their flux densities between the JVLA2011 sub-epochs and/or the JVLA2018 sub-epochs. It is worth noting that the JVLA2011 sub-epochs are more sensitive than the JVLA2018 sub-epochs. If we match the JVLA2011 sub-epochs detection threshold to the JVLA2018 sub-epoch detection thresholds, we find a higher variable fraction of $\sim2.5\%$ for the JVLA2011 sub-epochs. In total, we find that 7 sources are classed as short-term variables and only one these sources show variability between both sub-epochs. 

\begin{figure}
	\includegraphics[width=\linewidth]{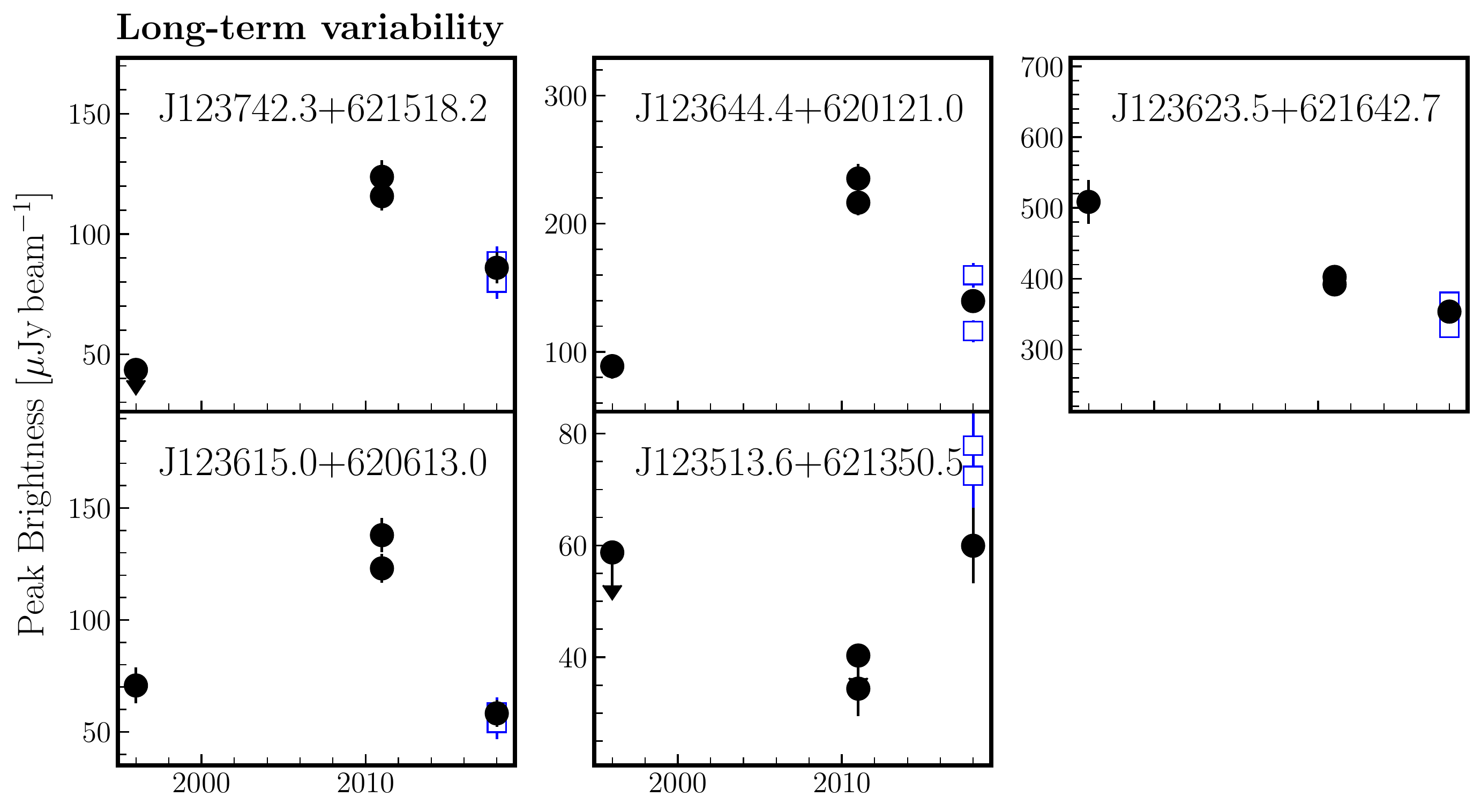}
	
	\includegraphics[width=\linewidth]{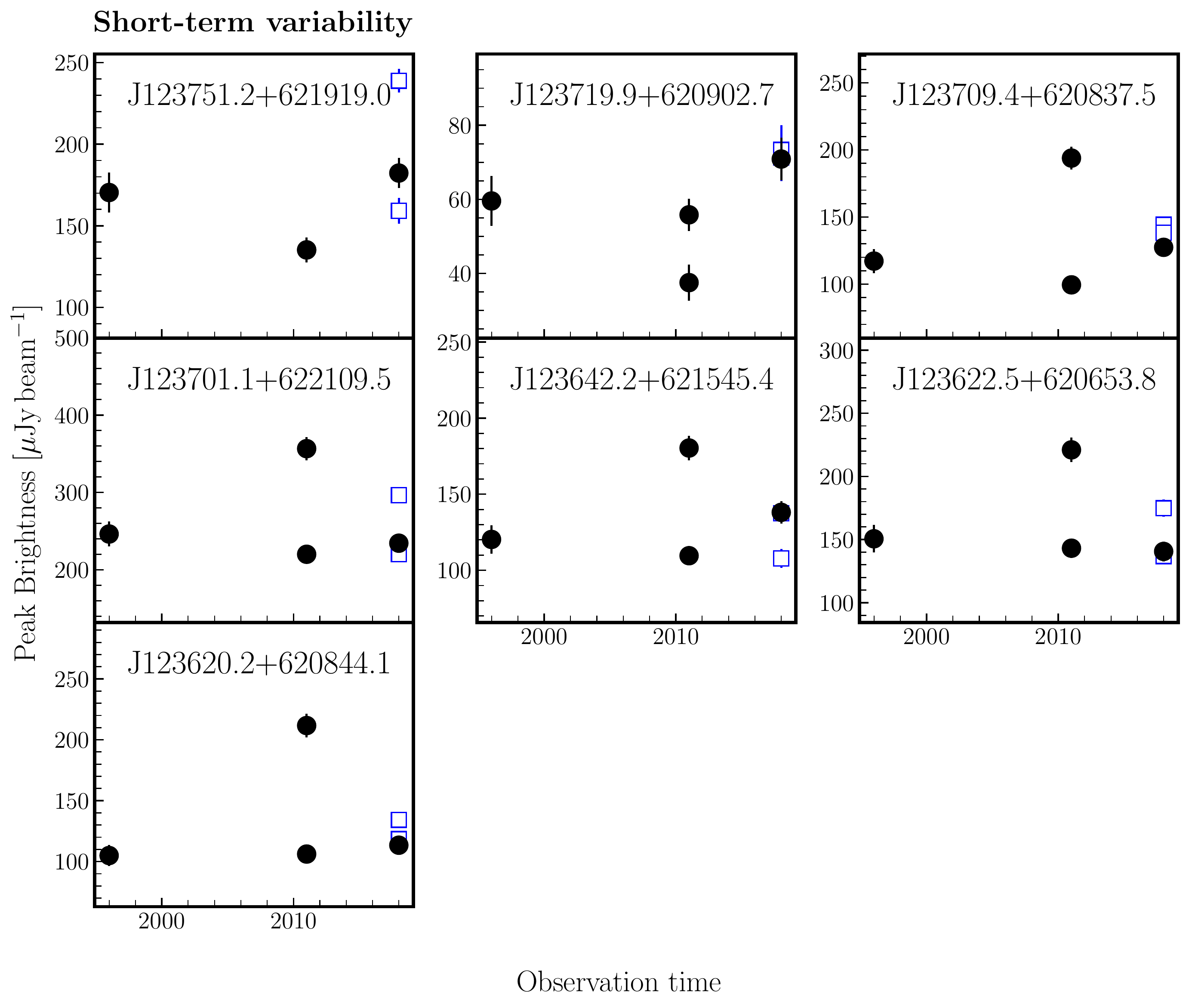}
	\caption{Light curves for the variable candidates. The top panel corresponds to those which are classified as long-term variable candidates and the bottom panel corresponds to short-term variable candidates. The entries with asterisks are both long and short cadence candidates. The open markers show the combined fluxes for the JVLA epochs illustrating how short-term variability can artificially induce long-cadence variability. }\label{Fig:light_curves}
\end{figure}

Based upon our definitions of long and short-term variability, we find that 5/10 variable sources are classed as both long and short term variables. In these sources, the short-term variability causes the flux to change sufficiently when the sub-epochs are combined such that the sources are also classed as long-term variables. We highlight this effect in the light curves presented in Figure~\ref{Fig:light_curves}, where the bottom panel shows the influence of the short term variability has upon the long term variability measurements. We can partially mitigate this in our sample by isolating those sources identified as long term variables only. This results in only three sources. However, it is worth stating that the sub-epoch cadence scales sampled here ($<$week) cannot rule out that the long term variability in this sample is being induced by $>$week timescale variability. We would expect that the physical mechanism causing variability on short timescales to be different to the longer timescales. For example, \citet{Ofek2011:var} suggest that most short term variability ($\sim$week timescales) is extrinsic to the source, whilst long term variability could be intrinsic (e.g. quiescent, AGN flaring). Therefore, in order to mitigate this properly and truly separate the physical mechanisms causing variability on different timescales, we would require multiple epochs which have cadences from day to decade timescales such that any short term variability that could induce long term variability can be identified.

To summarise, we find a total of 10 unique sources exhibiting variability which corresponds to $\sim 2\%$ of the total persistent radio source population comprising of 479 sources. Of these 10, we find that 8 are classed as long-term variables while 7 are short-term variables. There are 5 sources which exhibit both short and long term variability for which the flux-changes in the short-term influences the long-term variability. This leaves 3 long term variable candidates, however it is worth noting that these sources could still be influenced by $>$week scale variability.

This leads to a few conclusions. Firstly, the 1.4\,GHz long term variable radio sky is relatively sedate, continuing the trend of previous variability surveys with shorter cadences \citep[e.g.][]{Mooley2016:var,Hancock2016}. Secondly, at these flux densities, long cadence variables seem to contain less variable sources than the day to week timescale epochs with the fraction of purely long term variables being less than 1\% of the persistent radio population. However, we note that with the relatively small numbers of variable sources identified in this study, strong statistical conclusions cannot be made.

\subsection{Transient sources}

For the purposes of this paper, we define a transient source as a source which only appears in a single epoch and satisfies the variability criteria (based upon flux density upper limits derived from epochs where the source is not detected). We find no source which satisfies this criterion. We discuss the implications of this result in Section~\ref{SSec:transient_up_lims}. It is worth noting however, that by using this definition, a transient source could simply be a variable source at a flux density below the detection threshold which simply increases in flux during a single epoch. 
%

\section{Discussion}\label{Sec:discussion}
\subsection{Comparison to other variability surveys}

While this survey does not have the advantage of an ultra wide field of view which is advantageous for finding transients and variables \citep[see Appendix\,A of][]{Mooley2016:var}, it does have the considerable advantage of probing deeper than other surveys across extremely long decadal time scales. Other surveys at these frequencies suggests that the variable fraction is around 1\% or less on minute to year timescales \citep{Croft2010:var,Thyagarajan2011:var,Mooley2013:var,Bannister2011a:var}. However, deeper surveys suggest that the variable fraction is maybe higher. \citet{Carilli2003:var} studied variability in a similar manner to this study using a deep single VLA pointing at 1.4\,GHz in the Lockman Hole region. They derived an upper limit of $\sim 2\%$ of sources above a detection threshold of $100\,{\rm\mu Jy\,beam^{-1}}$ varying by more than 50\% over 19 days and 17 month cadences. This is largely consistent with the results presented here. In our short timescale study (3 day cadence) we find no variable sources above $100\,{\rm\mu Jy\,beam^{-1}}$ have a fractional variability >50\% ($|m|>0.667$). 

There is evidence that the number of variable sources detected is a function of sampling cadence. \citet{Ofek2011:var} carried out a 5\,GHz survey across 2.66 square degrees of the sky at low Galactic latitudes which aimed to investigate the variable population on short ($<$day) timescales. For sources above flux densities of 1.8\,mJy, they found a high fraction of persistent sources ($\sim 30\%$) were variable at the $>4\sigma$ confidence level across days to weeks cadences. This is significantly larger than the results in other surveys, which they attribute to other blind surveys being insensitive to fast variability (e.g. scintillation), caused by averaging across multiple observations. In addition, this survey was conducted at 5\,GHz, which has a selection bias towards flat-spectrum radio AGN. To this end, \citet{Ofek2011:var} presented a structure function of the variability (defined as $(S-\langle S\rangle)/\langle S\rangle$) which showed that the amount of variability is high on day to week timescales and is constant above around 10 days (albeit with no constraints on longer timescales). Our results support this, because we find that 70\% of our variable sources are variable the shortest timescales (days) which could correspond to the high levels of variability suggested by the \citet{Ofek2011:var} structure function. 

\citet{Mooley2016:var} extended upon this using the 3\,GHz CNSS 50 deg$^{2}$ pilot survey covering around 3500 radio sources. They find that 2.6\% of sources vary on 1.5 year timescales which is considerably more that the weekly (1\%) and monthly (0.8\%) cadences which is suggestive that the number of variable sources is a function of time up to the yearly cadences. In total, they find that $3.8\substack{+0.5 \\ -0.9}$\% of sources are variable on timescales $<1.5$ years. Ignoring the difference in frequency, our results for the shortest timescales (3 days) are consistent with $\sim 1.5\%$ of the sources showing variability. 

On the longest timescales, \citet{Hodge2013:var} compared the Stripe 82 VLA observation with FIRST and found that $\sim$ 6\% of sources in the mJy flux density regime show fractional variability above 30\% on 7-22 year timescales. This is considerably higher than seen in other studies which find that the fraction of variables is of the order a few percent or less \citep{Croft2010:var,Becker2010:var,Bannister2011b:var}. Our results presented here seem to agree with the latter studies as we find a long term variable fraction of $\leq1\%$.

\citet{Bannister2011a:var} found some evidence that there is a peak in radio variability on timescales of between 2000 and 3000 days. This was reinforced by \citet{Hancock2016} using a 1.4\,GHz study of variability in the Phoenix deep field. Our results suggest that variability reduces in the long decadal cadences and as such implies that the peak in variability suggested by \citet{Bannister2011a:var} and \citet{Hancock2016} may be viable. However, it is worth noting that this survey is probing a different flux density regime (0.03-2\,mJy) to the majority of surveys before (typically $>0.5$\,mJy). In the standard picture, the radio source population at the flux densities probed by this paper are transitioning from an AGN to a star-forming galaxy dominated regime, thus the smaller fraction of variable sources may also be an imprint of the change in the nature of the underlying source population.

\begin{landscape}
\begin{table}
	\centering
	\caption{A summary of the multi-wavelength properties and AGN classification for the variable candidates.  Long and short term variable candidates are denoted by L and S respectively in the Var. type column. Checkmarks or bold font indicates that the source is classed as an AGN. Redshifts with 68\% upper and lower bounds are photometric. The redshift reference abbreviations are as follows: W04 - \citet{Wirth2004:rs}, S04 - \citep{Smail2004:rs}, B08 - \citet{Barger2008:rs}, S14 - \citet{Skelton2014:rs}, Y14 - \citet{Yang2014:rs}. The star formation rates (SFR) were compiled from \citet{Whitaker2014:sfr} who used a combination of IR+UV to derive SFRs. The VLBI checkmarks in brackets correspond to those with $6\mbox{-}7\sigma$ VLBI counterparts coincident with the e-MERLIN positions and not reported in \citet{Chi2013:vlbi} or \citet{Radcliffe2018}. The Donley column corresponds to the IR-AGN classification technique proposed by \citet{donley2012identifying} and the WISE column corresponds to the \citet{Stern2012:ir} AGN classification technique. The crosses indicate that the sources had the bands necessary to evaluate these metrics but were not classified as AGN. Entries with hyphens indicate that the measurement could not be taken as the source was out of the field-of-view of the required bands while empty entries indicate that the source was not-detected in the required bands for classification.}
	\label{tab:multi_wave_properties}
\begin{tabular}{ccccccccccccc}
\hline
Source ID & Var. type &$z$ & $z$ ref. & Morph. & SFR & $\langle S_\mathrm{p}\rangle$ &  $q_{24}$ & Donley & WISE & X-rays & VLBI & e-MERLIN\\
& & && & [$\rm M_\odot\,yr^{-1}$]  & [$\rm\mu Jy\,beam^{-1}$] & & && [$\rm erg\,s^{-1}$] & & \\
\hline
J123742.33+621518.27 & L & $0.07$ & B08 & 2 & 0.4  & 103 & 0.92$\pm$0.04 & $\times$ &  & $\mathbf{2.3 \times 10^{40}}$ & (\checkmark) & \checkmark \\
J123623.55+621642.73 & L & $1.918$ & S04 & 1 &  & 419 & $\mathbf{-1.08\pm0.08}$ & $\times$ &  &  & \checkmark & \checkmark \\
J123615.02+620613.06 & L & $1.2\substack{+0.10 \\ -0.06}$ & Y14 & 1 & - & 86 & - &  &  &  &  & \checkmark \\
J123751.23+621919.00 & LS & $1.11\substack{+0.03 \\ -0.02}$ & S14 & 1 & 5.6 & 163 &  & $\times$ &  &  & \checkmark & \checkmark \\
J123709.43+620837.55 & LS & $0.907$ & W04 & 1 & 11.4 &  134 & $\mathbf{-0.57\pm0.14}$ & $\times$ &  & $\mathbf{3.2 \times 10^{42}}$ & \checkmark & \checkmark \\
J123644.48+620121.09 & LS & $0.53\substack{+0.03 \\ -0.03}$ & Y14 & 1 & - & 151 & -  & - & $\times$ & - &  & - \\
J123622.51+620653.90 & LS & $1.94\substack{+0.12 \\ -0.12}$ & S14 & 3 & 55.0 & 161.0 & $\mathbf{-0.3\pm0.05}$ & $\times$ &  &  & \checkmark & \checkmark \\
J123620.27+620844.15 & LS & $1.016$ & B08 & 1 & 3.5 &  122 &  & $\times$ & $\times$ &  & \checkmark & \checkmark \\
J123701.11+622109.55 & S & $0.8$ & B08 & 1 & 6.2 &  262 &  & $\times$ & $\times$ & $\mathbf{1.5 \times 10^{42}}$ & \checkmark &   \\
J123642.21+621545.42 & S & $0.858$ & B08 & 2 & 85.7 &  137 & $0.74\pm0.04$ & $\times$ & $\times$ & $\mathbf{9.9 \times 10^{42}}$ & \checkmark & \checkmark \\
\hline
\end{tabular}
\end{table}
\end{landscape}

\subsection{Sources of variability}\label{SSec:variability_sources}

To understand the nature of the variability in these sources, we compiled multi-wavelength data for each of these sources utilising the extensive photometry of objects in the GOODS-N field. These data are summarised in Table~\ref{tab:multi_wave_properties}. Redshifts were compiled for the 10 sources from various catalogues \citep{Wirth2004:rs,Smail2004:rs,Barger2008:rs,Skelton2014:rs,Yang2014:rs,Momcheva2016:rs}. Of these 12 redshifts, 6/10 are spectroscopic redshifts and the rest are photometric redshifts (of which one is uncertain). The redshifts range from 0.07 to 1.94 with a median redshift of 0.96. 

Optical and near infra-red (NIR) Hubble Space Telescope (HST) data \citep{Giavalisco2004:hst,Skelton2014:rs} plus CFHT $K_s$ band imaging \citep{Wang2010:ir} were used to determine the host galaxy morphologies. For the morphologies, we visually classified the sources into four categories: 1 - Early-type / bulge dominated, 2 - late-type / spiral galaxies, 3 - irregular or 4 - unclassified. 

The early type / bulge dominated group are those circular/elliptical extended objects whose surface brightness distribution drops towards the edge, while late type galaxies must have clear spiral features. The irregular category encompasses those with clumpy surface brightness distributions and sources which are unclassified are those for which a morphology cannot be attained due to being faint. These sources often have undetected low surface brightness areas hence they often appear `point-like'. 

We find that the majority of all variable sources (70\%) have early-type morphologies which are typical of radio-loud AGN \citep[e.g.][]{Hickox2009,Griffith2010}. The remainder comprise of two late-type galaxies and one irregular galaxy. To establish whether the radio emission is constrained to the nuclear region we compared these optical images to the VLA, e-MERLIN (Muxlow et al. in prep.) and VLBI positions \citep{Chi2013:vlbi, Radcliffe2018}. Note that the e-MERLIN data forms part of the upcoming e-MERGE survey (Muxlow et al. in prep.; Thomson et al. in prep.)

The VLBI observations alone provide the most compelling evidence that the radio emission is caused by an AGN. For extragalactic sources, a VLBI detection implies brightness temperatures in excess of $10^5\,\mathrm{K}$ which cannot be reliably attributed to star-formation processes alone and must instead be AGN related \citep{Condon1992:rad,Kewley2001:vlbi}. We find that 80\% (8/10) of the variable candidates exhibit VLBI emission. This includes one source with a 6-7$\sigma$ VLBI detection coincident with the e-MERLIN emission, which was not formally reported in \citet{Radcliffe2018} due to their adopted 7$\sigma$ detection threshold. Interestingly, J123642+621545 was measured to have a total flux density of 343 $\mathrm{\mu Jy}$ in the 2004 Global VLBI observations of \citet{Chi2013:vlbi}, almost 2.5$\times$ the VLA flux density. This has subsequently reduced in flux density significantly and was not detected in the 2014 EVN observations of \citet{Radcliffe2018}. All of the VLBI positions acquired are consistent with nuclear activity, apart from J123742.33+621518.30 which seems to be offset by around 0\farcs24 from the optical maximum. We discuss this source in Section~\ref{SSec:possible_SNII}. 

For the remaining 2 sources which are not detected by VLBI observations, we use the more sensitive e-MERLIN observations ($\sim3\mathrm{\mu Jy\,beam^{-1}}$; Muxlow et al. in prep.) to search for evidence of nuclear emission in the remaining sources. This resulted in one more source with radio emission coincident with the nuclear regions while the remaining source was out of the field-of-view of the e-MERLIN observations.


We searched for other indications of AGN activity in these sources in the infra-red (IR) and X-ray bands. Additional \textit{Spitzer} IRAC \citep[3.6-8$\rm\,\mu m$][]{Wang2010:ir, ashby2013:ir, Yang2014:rs} and MIPS \citep[24, 70$\rm\,\mu m$][]{dickinson2003spitzer, Magnelli:2011eb} plus \textit{WISE} \citep[3-22$\rm\,\mu m$,][]{Cutri2012:WISE} photometry was compiled. The X-ray data was derived from the 0.5-7\,keV, $\mathrm{2\,Ms}$ \textit{Chandra} exposure with the catalogue provided by \citet{xue_xray_2016}. A 1\arcsec~search radius was used and any counterparts were visually evaluated using the higher resolution radio and optical HST images. 

\begin{figure}
	\includegraphics[width=1.0\linewidth]{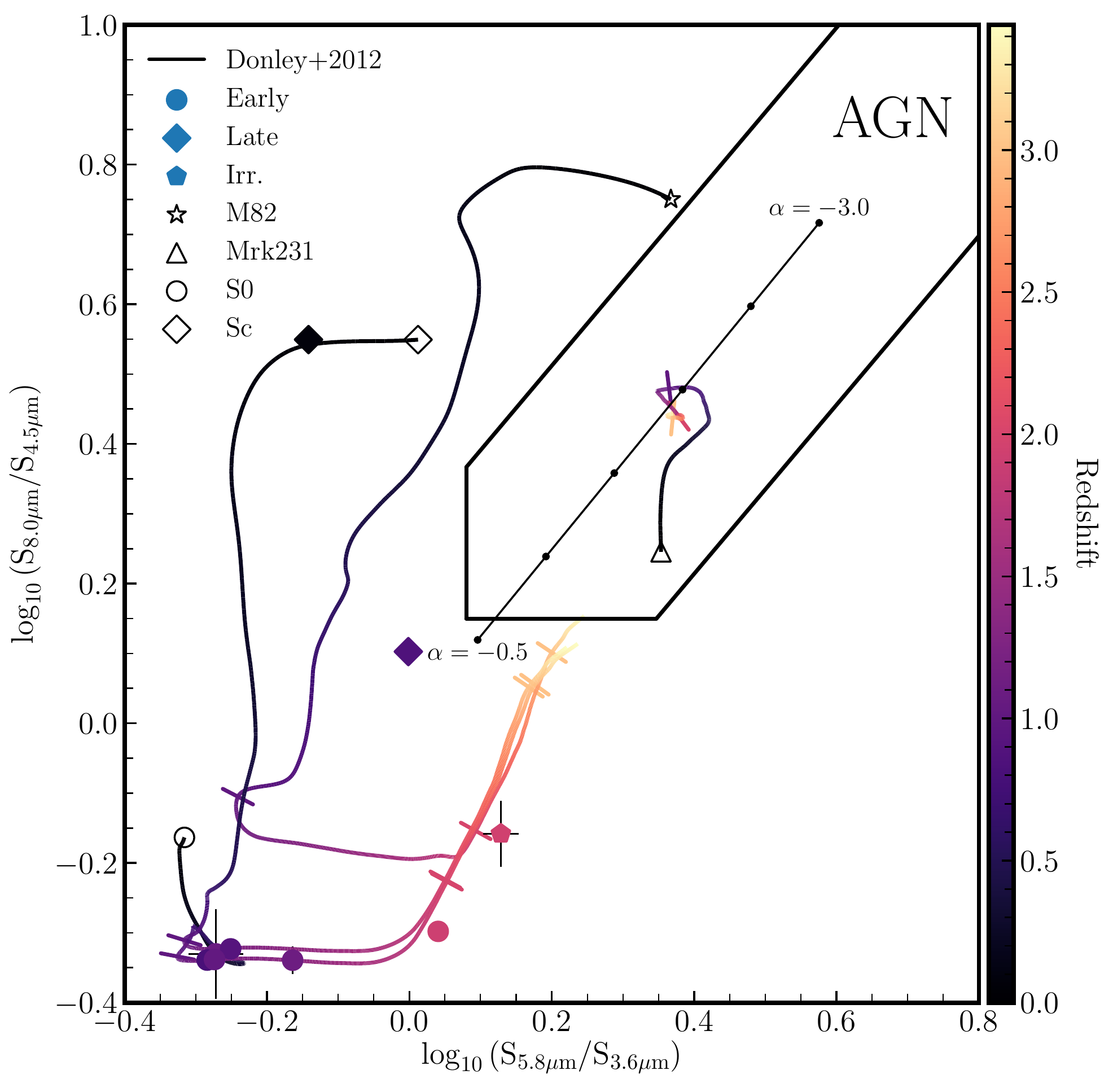} 
	\caption{\,\textit{Spitzer} IRAC AGN selection criteria for the variable candidate sample. The solid line corresponds to the revised AGN selection wedge by \citet{donley2012identifying}. The solid line with black markers corresponds to the IR power law locus in the range $ -3.0\leq\alpha\leq-0.5$. Overlaid are the predicted SED colours of the starburst galaxy M82, AGN Mrk231, an Sc type and an S0 type galaxies from the SWIRE library \citep{Polletta2006:sed}. The tracks are color-coded across a redshift range of 0-3.4 with the perpendicular bars corresponding to integer redshift intervals. The marker colors of the variable sample correspond to their redshift and are matched to the track colors while the marker shapes correspond to the host morphology. The AGN selected objects using these criteria are those in the top-right of the figure for all selection methods}\label{Fig:IR_AGN}
\end{figure}

These multi-wavelength data can be used to further distinguish between AGN or star-formation related emission mechanisms. For {\it Spitzer} IRAC NIR bands, we used the IR colour-colour diagnostics of \citet{donley2012identifying} to establish whether the source had AGN signatures. Here the presence of a dusty AGN torus causes the spectral energy distribution (SED) to represent a power law in the region between the $1.6\mathrm{\mu m }$ stellar bump and the 25-50\,K emission from cold dust heated by star-formation. The \citet{donley2012identifying} criterion classifies a source as an AGN if it satisfies all of the following:
	
\begin{align}
x &\geq 0.08 \nonumber\\
y &\geq 0.15 \nonumber\\ 
y &\geq(1.21 \times x)-0.27 \nonumber\\ 
y &\leq(1.21 \times x)+0.27 \nonumber\\ 
S_{4.5 \mu \mathrm{m}}&>S_{3.6 \mu \mathrm{m}} \land S_{5.8 \mu \mathrm{m}}>S_{4.5 \mu \mathrm{m}} \land S_{8.0 \mathrm{m}}>S_{5.8 \mathrm{um}}\nonumber
\end{align}

\noindent where $x = \log_{10}(S_\mathrm{5.8\mu m}/S_\mathrm{3.6\mu m})$ and $y=\log_{10}(S_\mathrm{8.0\mu m}/S_\mathrm{4.5\mu m})$. In total we found that 8/10 sources have NIR counterparts in the four \textit{Spitzer} IRAC bands with S/N ratios larger than $3\sigma$ in each band. As Fig~\ref{Fig:IR_AGN} shows, the majority of these sources show little or no signs of AGN activity in the infra-red with their NIR characteristics closely following star-formation dominated emission, as shown by the close correlation between the host morphology and the observed NIR colours. For those sources outside of the \textit{Spitzer} IRAC coverage, we cross matched to the WISE all-sky catalogue \citep{Cutri2012:WISE} which yielded one additional match. Due to the poor sensitivity in the W3 and W4 bands (only one detection in band 3 and none in band 4), we used the \citet{Stern2012:ir} criterion ($\rm W1-W2\leq0.8$) in order to identify AGN. Again, this source showed no IR-AGN signatures.

In addition, we use the \textit{Spitzer} MIPS 24$\rm\mu m$ flux densities to identify those sources which deviate from the IR-radio correlation. Despite there being some contamination effects \citep[e.g. from high redshift AGN,][]{DelMoro13:agn}, the ratio between the 24$\rm\mu m$, $S_\mathrm{24\micron}$ and 1.4\,GHz flux densities, $S_\mathrm{1.4\,GHz}$ can be used effectively distinguish between AGN and star-formation related activity where systems with an AGN present will produce more radio emission than what is expected from star-formation alone \citep[e.g.][]{Appleton2004:ir,Garn2009:qir,Chi2013:vlbi}. In addition, this band was chosen to alleviate the effects of confusion which plagues the longer wavelength observations. For each source with $24\mu m$ counterparts (5/10), we calculated $q_{24}$ using,

\begin{equation}
q_{24} = \log_{10}(S_{\rm 24\mu m}/\langle S_{\rm1.4\,GHz}\rangle).
\end{equation}

\noindent For those sources with $24\rm\mu m$ counterparts, we find that three show clear radio-excess signatures from AGN activity \citep[$q_{24}<0$;][]{DelMoro13:agn}. We also cross-matched the sample with the 2Ms \textit{Chandra} X-ray observations presented in \citet{xue_xray_2016}. In total we find only 4 sources have X-ray counterparts all of which are classified as AGN by their selection criteria \citep[see Section 2.3.5 of][]{xue_xray_2016}. 

In light of this, these commonly used multi-wavelength diagnostics are not able to identify all these variable sources as AGN and three sources rely on the identification of a compact radio component. Indeed based upon the star formation rate (SFR) measurements and IR colours, some of these systems (e.g. J123642.21+621545.42) only show evidence of AGN activity by the existence of a high brightness temperature core revealed by the VLBI and e-MERLIN observations. This reinforces our belief that high resolution radio observations are integral to obtaining a full consensus of AGN activity across cosmic time. 

In total, we find compelling evidence for AGN activity in 8/10 sources. Of the remaining sources, J123742.33+621518.27 is a possible supernovae (see Sec.\,\ref{SSec:possible_SNII}) whilst J123644.48+620121.08 is outside of the e-MERLIN and VLBI field-of-view. However it is worth noting that it is hosted by a typical QSO host, an elliptical galaxy.

Almost all of the variable objects are detected with VLBI, illustrating that there is a milliarcsecond scale compact radio source producing the variability seen here. Recent VLBI surveys have shown that VLBI-detected sources are becoming more compact at lower flux densities \citep{DellerMiddelberg2014:vlbi,Radcliffe2018}. This is supported by semi-empirical simulations which require FR-I type sources to become core-dominated at low flux densities in order to reconcile the high-frequency ($>10\,\mathrm{GHz}$) number counts with the low-frequency number counts ($<5\mathrm{GHz}$) \citep{Whittam2017}. It could be that the variable population at low flux densities traces this population because variable sources will inherently contain a compact radio component. Indeed, the lack of mid-IR AGN signatures (which require a dusty AGN torus to re-radiate UV photons into the mid-IR bands) in any of these objects further adds to this argument, as FR-I galaxies are known to not have dusty tori \citep[e.g.][]{Volk2010}.

For the short term variable objects, the physical origin of the variability could be scintillation. This has been previously been suggested as the primary cause of most short-term variability at frequencies below 5\,GHz.
\citep[e.g.][]{Qian1995,Frail2000,Ofek2011:var}. For the long-term variable sample, the source of variability is most likely to be intrinsic to the source and could be caused by changes in the AGN jet power which has timescales of years to decades \citep[e.g.][]{Aller1999,Arshakian2012:vlbi} or shocks in the jet \citep[e.g.][]{WooUrry2002,Hovatta2008}. It is worth noting that, without continual monitoring of the source, the origin of the variable emission cannot be distinguished easily. 
\begin{figure*}
	\includegraphics[width=\linewidth,clip]{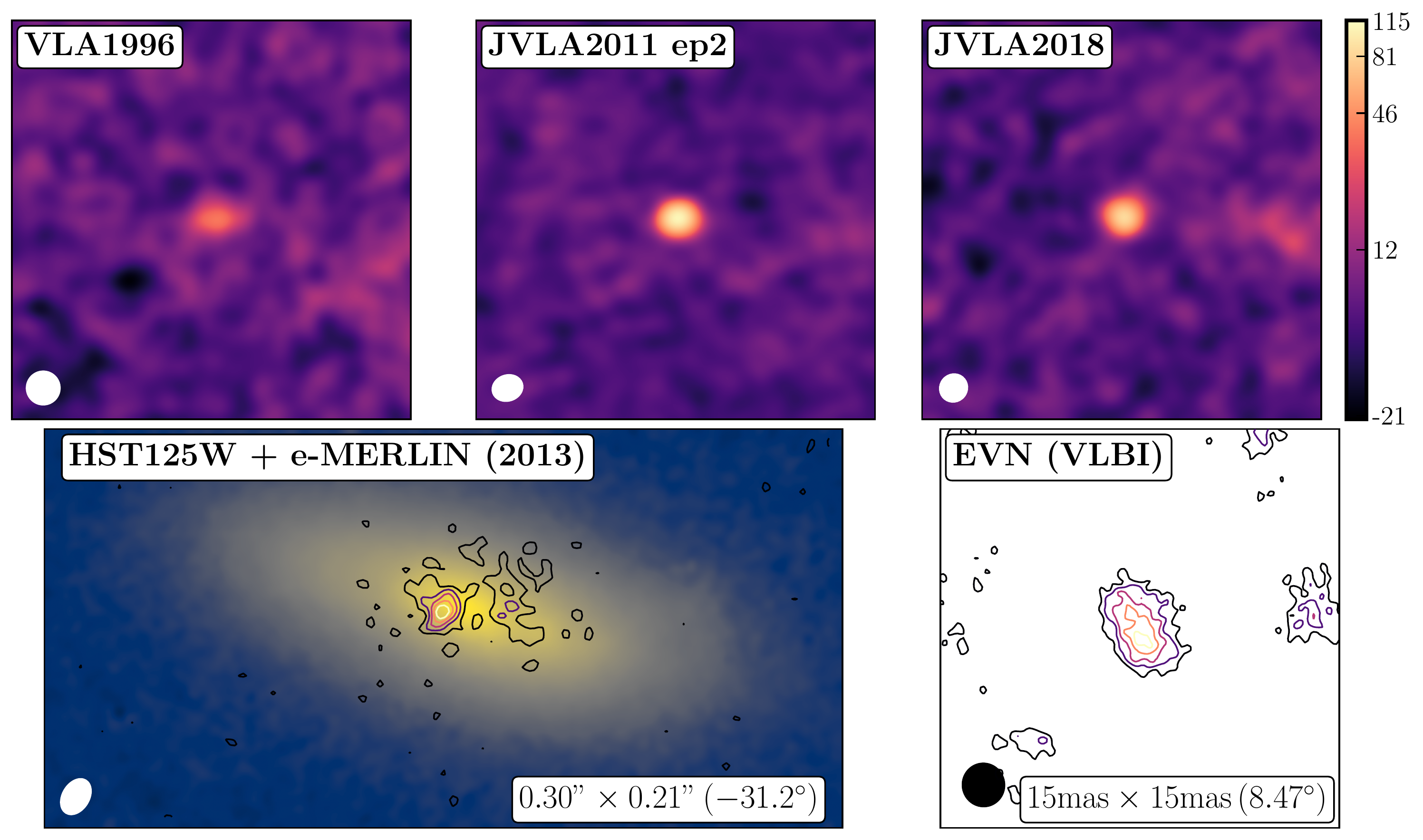} 
	\caption{Type II supernovae / LLAGN candidate. The top row shows the evolution of this source from the VLA1996 to the JVLA2018 epochs. The colour scale is identical for each VLA epoch and are in units of $\mathrm{\mu Jy\,beam^{-1}}$. The bottom row illustrates the e-MERLIN offset (Muxlow et al. in prep.) from the astrometry corrected HST125W optical image \citep{Grogin2011,Koekemoer2011,Skelton2014:rs} with the 6.7$\sigma$ VLBI detection in the bottom right panel. }\label{Fig:sn_candidate}
\end{figure*}
For a sparsely sampled cadence of observations as presented here, the physical processes which cause variability on month to year timescales can masquerade as processes on much longer timescales. To solve this, more epochs spread over short and long cadences are needed to distinguish between the two.

\subsection{J123742.33+621518.27 - a possible radio supernova?}\label{SSec:possible_SNII}

This source was noted to be unusual due to the offset between the radio position (provided by e-MERLIN) and optical emission ($\sim0\farcs24$) along with low redshift ($z=0.07$) of the host galaxy. As shown in Figure~\ref{Fig:sn_candidate}, the VLA1996 data does not show the compact component which has appeared in the JVLA2011 epoch. The peak radio luminosity from the 2012 JVLA 5.5\,GHz observations of \citet{Guidetti2017:5} is $5.67\times10^{20}\,\mathrm{W\,Hz^{-1}}$ and the spectral index of $-0.71$. This is a typical luminosity for core-collapse radio supernovae, which can range between $5\times10^{19}$ and $1.3\times10^{22}\,\mathrm{W\,Hz^{-1}}$ \citep{Weiler2003:SN}. The $q_{24}$ radio excess measure shows no radio-excess emission which would be indicative of an AGN and there are no signs of AGN activity in the IR diagnostics. 

Prior to the emergence of this new compact component the VLA1996 observations detected diffuse radio emission which was below the formal 7$\sigma$ detection threshold used in this variability study. We hypothesis that this diffuse off-nuclear emission may originate from on going star-formation processes in the galaxy hence the diffuse radio emitting HII region. Low resolution WSRT imaging \citep[observed in May 1999;][]{Garrett2000} recorded an integrated flux density of $\sim 180\,{\rm\mu Jy}$, which is far in excess of the flux densities recorded by the later JVLA2011 and 2018 epochs (see Figure~\ref{Fig:light_curves}). These JVLA epochs show that a new compact radio component dominates over the diffuse emission seen in the VLA1996 epoch. e-MERLIN imaging (observed in 2013) (Figure.\,\ref{Fig:sn_candidate}) shows this diffuse emission across the face of the optical host along with an offset compact component. This diffuse emission could be the same star-forming region detected in the VLA1996 data or could be from a low-luminosity AGN jet. The compact source has a 6.7$\sigma$ EVN detection (observed in 2014) which corresponds to an estimated brightness temperature of $\sim1.8\times 10^{5}\,\mathrm{K}$. This is slightly larger than expected from star-formation processes alone \citep{Condon1991}.

The physical origins of this off-nuclear compact radio source remain unclear but the luminosity, spectral index and light-curve timescales are broadly consistent with the characteristics of a powerful core-collapse radio supernovae which has occurred between 1996 and 1999 which now dominates the radio emission from the HII region. If this new variable source is a core-collapse radio supernovae, it would be the most distant ever discovered.

Unfortunately, with the information available, it is not possible to categorically determine if this off-nuclear variable source is a distant radio supernovae or a variable off-nuclear accretion powered source such as one component of a wide separation binary AGN system. However this remains an intriguing variable source which will be monitored by future VLBI observations to investigate any expansion of the compact radio emitting region.


\subsection{Transient upper limits}\label{SSec:transient_up_lims}

\begin{figure}
	\includegraphics[width=0.95\linewidth]{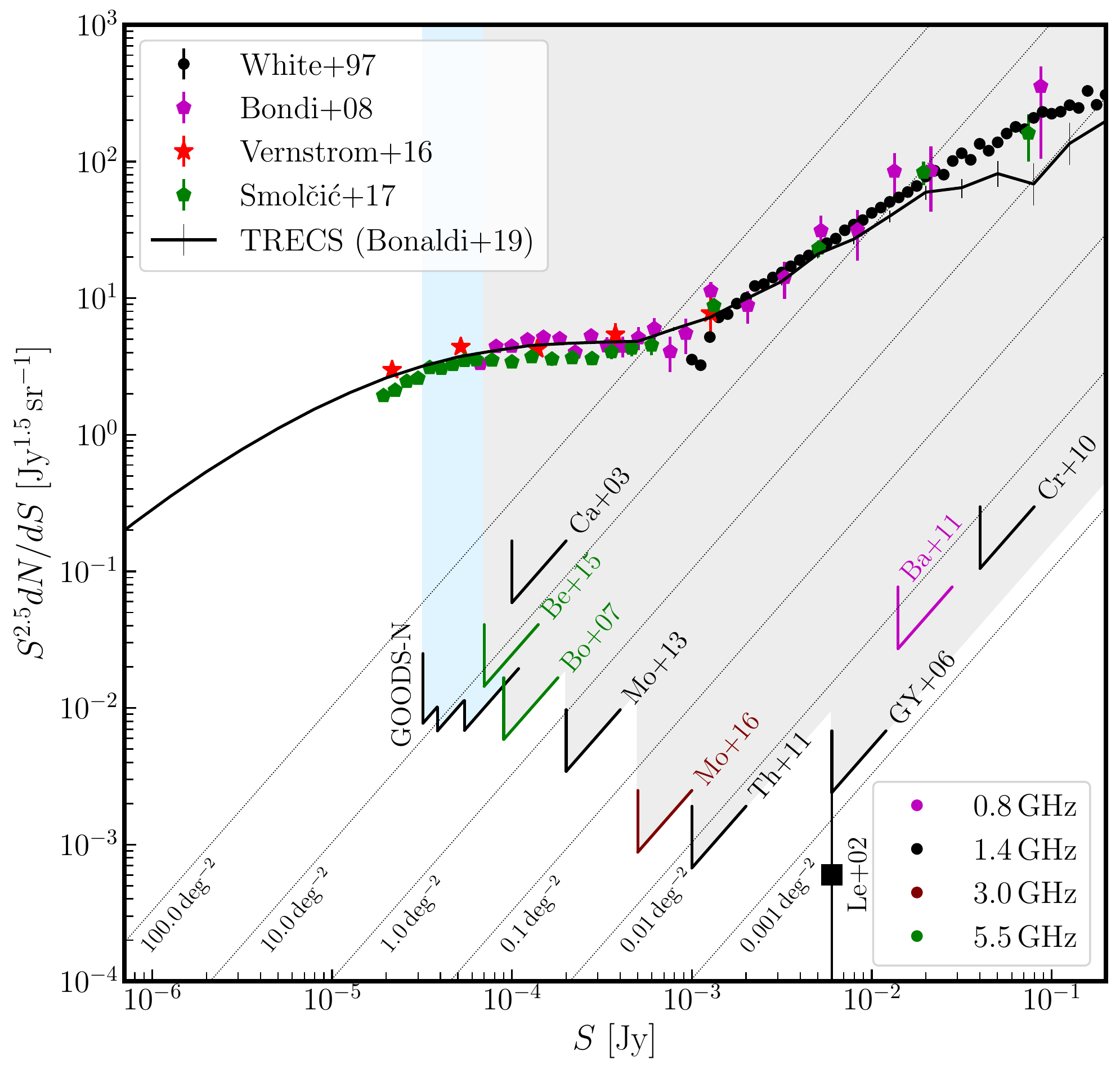} 
	\caption{The transient normalised areal densities and upper limits with respect to flux density. For comparison the normalised differential number counts for persistent sources are overlaid \citep{White1997:first,Bondi2008:vla,Vernstrom2016:con,Smolcic2017:sc}. Over-plotted are the expected total extragalactic differential number counts from the TRECS simulation \citep{bonaldi2019:trecs}. The wedges correspond to the various 95\% upper limits for surveys without transient detection and their color corresponds to the survey frequency. The shaded area shows the phase space excluded by the current surveys and the abbreviations corresponds to the following references: Ca+03 - \citet{Carilli2003:var}, Bo+07 - \citet{Bower2007}, Be+15 - \citet{Bell2015} , Mo+13 - \citet{Mooley2013:var}, Mo+16 - \citet{Mooley2016:var}, GY+06 - \citet{GY2006}, Ba+11 - \citet{Bannister2011a:var}, Cr+10 - \citet{Croft2010:var}, Le+02 - \citet{Levinson2002}.}\label{Fig:transient_constraints}
\end{figure}

In this survey, we detected no transient sources i.e. those sources which only appear in one epoch alone. Despite the small FoV and therefore limiting constraints that this places on the transient fraction we estimated an upper limit to the transient fraction was calculated following the prescription of \citet{Ofek2011:var}. If we assume an arbitrary luminosity function, a uniform density of transient sources in a Euclidean universe and a primary beam that can be reliably approximated by a Gaussian, then the surface density, $\kappa$, of transients brighter than flux density, $S$, can be calculated using, 

\begin{align}\label{eqn:transient_rates}
	\kappa(>S) =&  \kappa_0\left(S/S _ { \min , 0 }\right) ^ { -3 / 2 } ,\\
	\kappa_0 =& \frac { 3 N _ { b } \ln 2 } { 2 \pi r _ { \mathrm { HP } } ^ { 2 } }\left( 1 - \exp\left( - 3 r _ { \max } ^ { 2 } \ln(2) /  2 r _ { \mathrm { HP } } ^ { 2 } \right) \right) ^ { - 1 }
\end{align}

\noindent where $N _ { b }$ is the number of transients, $r _ { \mathrm { HP } }$ is the primary beam HPBW, $r_\mathrm{max}$ is the maximum radius considered, $S_\mathrm{min,0}$ is the detection limit at $r=0$ i.e. the primary beam maxima. For the case considered here (no transients), the $2\sigma$ limit is three events \citep[$N_b=3$,][]{Gehrels1986}. The tightest constraints require the largest number of independent epochs. In this condition the limiting detection flux density is constrained by the JVLA2018 sub-epochs which has a central $7\sigma$ detection limit of 54.4\,$\mathrm{\mu Jy\,beam^{-1}}$. The primary beam has a $r_\mathrm{HP} = 0.258\,\mathrm{deg}$ and transients were searched to a maximum radius of $r_\mathrm{max} = 0.233\,\mathrm{deg}$. Substitution into equation~\ref{eqn:transient_rates} and dividing by the number of independent epochs (5) gives a 2$\sigma$ upper limit on the areal density $\kappa(>54.4\microJybm) = 5.21\,\deg^{-2}$. This can be repeated to lower flux densities using different independent epoch combinations. These new constraints are shown Figure~\ref{Fig:transient_constraints} which shows the differential source counts and the limits excluded by previous radio surveys and the results shown here. We have included the differential source counts of persistent sources from observational surveys \citep{White1997:first,Bondi2008:vla,Vernstrom2016:con,Smolcic2017:sc} and the Tiered Radio Extragalactic Continuum Simulation \citep[TRECS;][]{bonaldi2019:trecs} for comparison. Our results illustrate that there is no upturn in transient events in the $\microJy$ flux density regime. This is entirely expected, however it is worth stating here that the majority of these surveys have very different cadence scales, therefore each survey will have a different sensitivity to very-short lived transient events. One conclusion that could be made is that there is no significant population below our detection threshold that are extremely variable on long timescales. This follows onto the generally accepted paradigm that the $\rm\mu Jy$ radio source population becomes increasingly dominated by star-forming galaxies which would not be expected to exhibit strong variability on decadal timescales. 

\section{Conclusions}\label{Sec:conclusion}

We have conducted a study on the variability in the GOODS-N field using 5 epochs of JVLA and VLA data spread across 22 years. We find a total of 10 variable sources, of which 8 show variability on long (year-decade) timescales while 7 show variability on very short (day) timescales. However, we find that the variability in five of the long term variables is maybe influenced by short term variability, leaving only 3 sources exhibiting true long-term variability. Nearly all variable sources (9/10) have peak brightnesses above 100\,$\rm\mu Jy\,beam^{-1}$, which could be an imprint of transition in the radio populations from AGN to star-forming galaxies at this flux density regime. 

Multi-wavelength and ultra-high resolution radio observations reveal that the variability in almost all of these sources can be reliably attributed to AGN activity. The lack of IR-AGN signatures, suggests that the variable source population in this regime could be the population of core-dominated FR-I galaxies, without dusty tori, that have been postulated in other wide-field surveys and semi-empirical simulations \citep{DellerMiddelberg2014:vlbi,Whittam2017}. The only source without an AGN signature may be a Type-II supernovae which, at $z=0.07$, would make it one of the most distant ever discovered. 

In light of this, transient and variable studies are a function of many different parameters such as cadence time, flux density, observing frequency, observing time and resolution. This paper highlights how important the range of cadences have upon the interpretation of the result. In the future, the separation of variability as a function of the aforementioned parameters will require multi-epoch wide-field observations. While such surveys have been conducted at the mJy and Jy regime \citep[e.g.][]{Bower2010,Croft2013,Mooley2016:var}, characterising the $\rm\mu Jy$ regime requires improved snapshot sensitivity and $uv$ coverage which will be provided by the next generation of interferometers, such as ASKAP, MeerKAT, SKA and ngVLA, and their associated transient projects such as VAST \citep{2013PASA...30....6M} and ThunderKAT \citep{Fender2017}.

\section*{Acknowledgements}
The authors would like to acknowledge the referee Jim Condon and the scientific editor for their useful comments that have greatly helped focus this paper. J.F.R. acknowledges support by the Science and Technologies Facilities Council (STFC), and the Ubbo Emmius Scholarship of the University of Groningen and the South African Radio Astronomy Observatory (SARAO) fellowship that funded this research. A.P.T. acknowledges support from STFC grant (ST/P000649/1).
The National Radio Astronomy Observatory is a facility of the National Science Foundation operated under cooperative agreement by Associated Universities, Inc. 
e-MERLIN is a National Facility operated by the University of Manchester at Jodrell Bank Observatory on behalf of STFC.
The European VLBI Network is a joint facility of independent European, African, Asian, and North American radio astronomy institutes. Scientific results from data presented in this publication are derived from the following EVN project codes: EG078, GG053. 



\bibliographystyle{mnras}
\bibliography{uJy_variability} 




\bsp	
\label{lastpage}
\end{document}